\documentclass[conference]{IEEEtran}

\usepackage{amsmath,amssymb,amsfonts}
\usepackage[noend]{algorithm}
\usepackage[noend]{algpseudocode}
\usepackage{graphicx,tabularx}
\usepackage{textcomp}
\usepackage{threeparttable}
\usepackage{color}
\usepackage{hyperref} 
\usepackage{subcaption}
\usepackage{enumitem}
\usepackage{verbatim}
\usepackage{booktabs}

\usepackage{array}
\newcolumntype{M}[1]{>{\centering\arraybackslash}m{#1}}

\newcommand{\xiao}[1]{\textcolor{blue}{#1}}

\newcommand{\cc}[1]{\textcolor{black}{#1}}

\begin{document}

\title{Low-Latency Video Conferencing via Optimized Packet Routing and Reordering}

\author{\IEEEauthorblockN{Yao Xiao\IEEEauthorrefmark{4}\IEEEauthorrefmark{5},
		Sitian Chen\IEEEauthorrefmark{2},
		Amelie Chi Zhou\IEEEauthorrefmark{2}\IEEEauthorrefmark{1},
		Shuhao Zhang\IEEEauthorrefmark{3},
		Yi Wang\IEEEauthorrefmark{4}\IEEEauthorrefmark{5}, 
		Rui Mao\IEEEauthorrefmark{4}\IEEEauthorrefmark{5} and
		Xuan Yang\IEEEauthorrefmark{4}\IEEEauthorrefmark{5}
	}
	\IEEEauthorblockA{\IEEEauthorrefmark{4}Shenzhen University \quad \IEEEauthorrefmark{2}Hong Kong Baptist University }	
	\IEEEauthorblockA{\IEEEauthorrefmark{3}Nanyang Technological University}
	\IEEEauthorblockA{\IEEEauthorrefmark{5}Guangdong Provincial Key Laboratory of Popular High Performance Computers}
}
\maketitle 

\begin{abstract}
In the face of rising global demand for video meetings, managing traffic across geographically distributed (geo-distributed) data centers presents a significant challenge due to the dynamic and limited nature of inter-DC network performance. Facing these issues, this paper introduces two novel techniques, \textit{VCRoute} and \textit{WMJitter}, to optimize the performance of geo-distributed video conferencing systems. VCRoute is a routing method designed for audio data packets of video conferences. It treats the routing problem as a Multi-Armed Bandit issue, and utilizes a tailored Thompson Sampling algorithm for resolution. Unlike traditional approaches, VCRoute considers transmitting latency and its variance simultaneously by using Thompson Sampling algorithm, which leads to effective end-to-end latency optimization.
In conjunction with VCRoute, we present WMJitter, a watermark-based mechanism for managing network jitter, which can further reduce the end-to-end delay and keep an improved balance between latency and loss rate.
Evaluations based on real geo-distributed network performance demonstrate the effectiveness and scalability of VCRoute and WMJitter, offering robust solutions for optimizing video conferencing systems in geo-distributed settings.

\end{abstract}

\IEEEpeerreviewmaketitle

\section{Introduction}
Due to the outbreak of the COVID-19 pandemic, the need for video conferencing increases dramatically throughout the world. According to Zoom statistics, the annual Zoom meeting minutes have increased by 3300\% from October 2019 to October 2020 and the number continues to grow~\cite{zoomstat}.
With the large number of businesses requiring global cooperation and communication, video conferencing systems such as Zoom and VooV have been commonly used in a geographically distributed (geo-distributed) manner, where global participants join an online meeting session from local clients to send and receive audio and video messages. Users' Quality of Experience (QoE) is heavily dependent on how well video conferencing systems transfer and process the messages between clients.

\begin{figure}[t]
    \centering
      \begin{subfigure}{0.4\linewidth}
 	\includegraphics[width=0.7\linewidth]{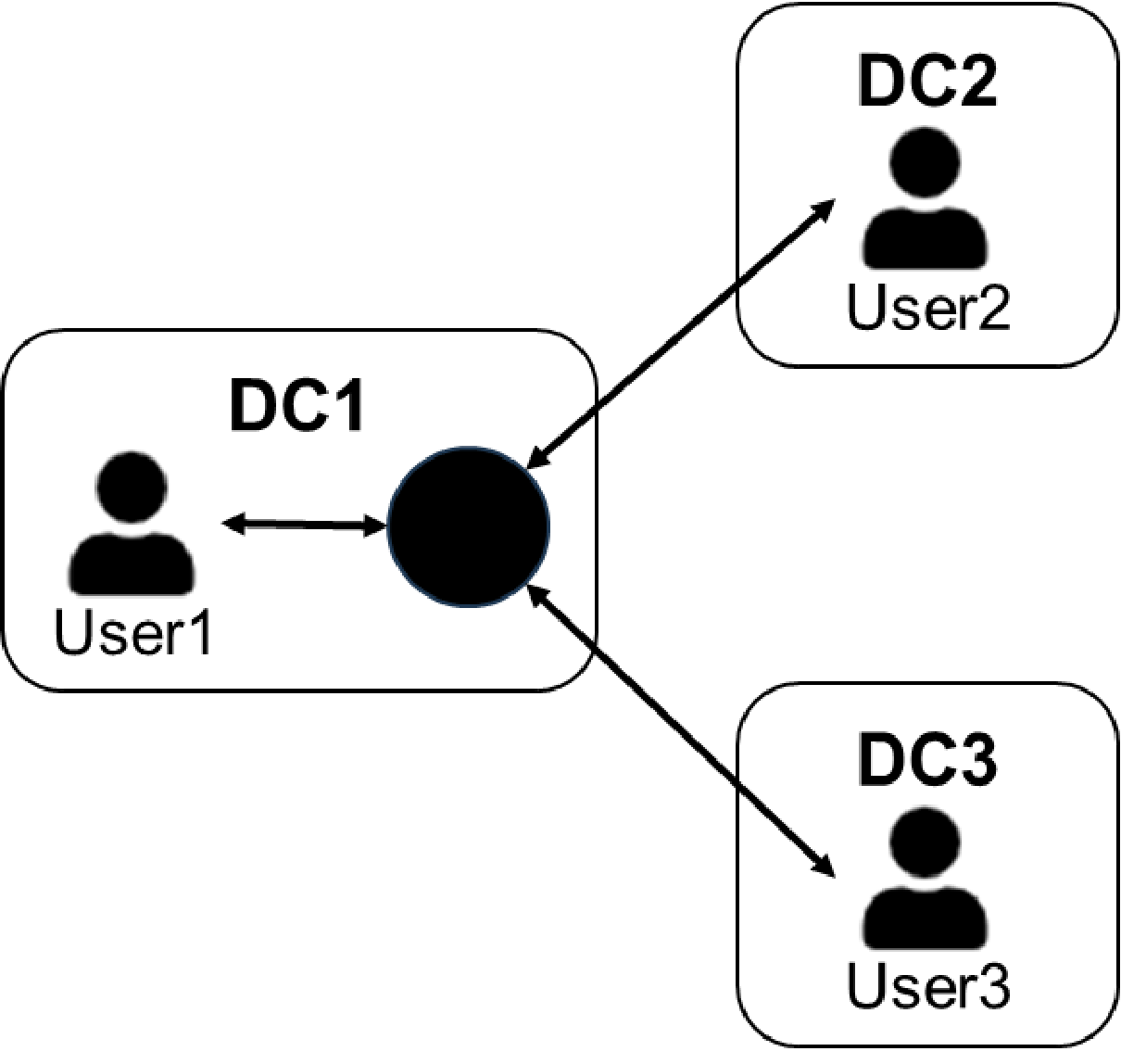}\caption{Single service endpoint}\label{fig:archit:a}
	\end{subfigure}
 \hfil
   \begin{subfigure}{0.5\linewidth}
 	\includegraphics[width=0.7\linewidth]{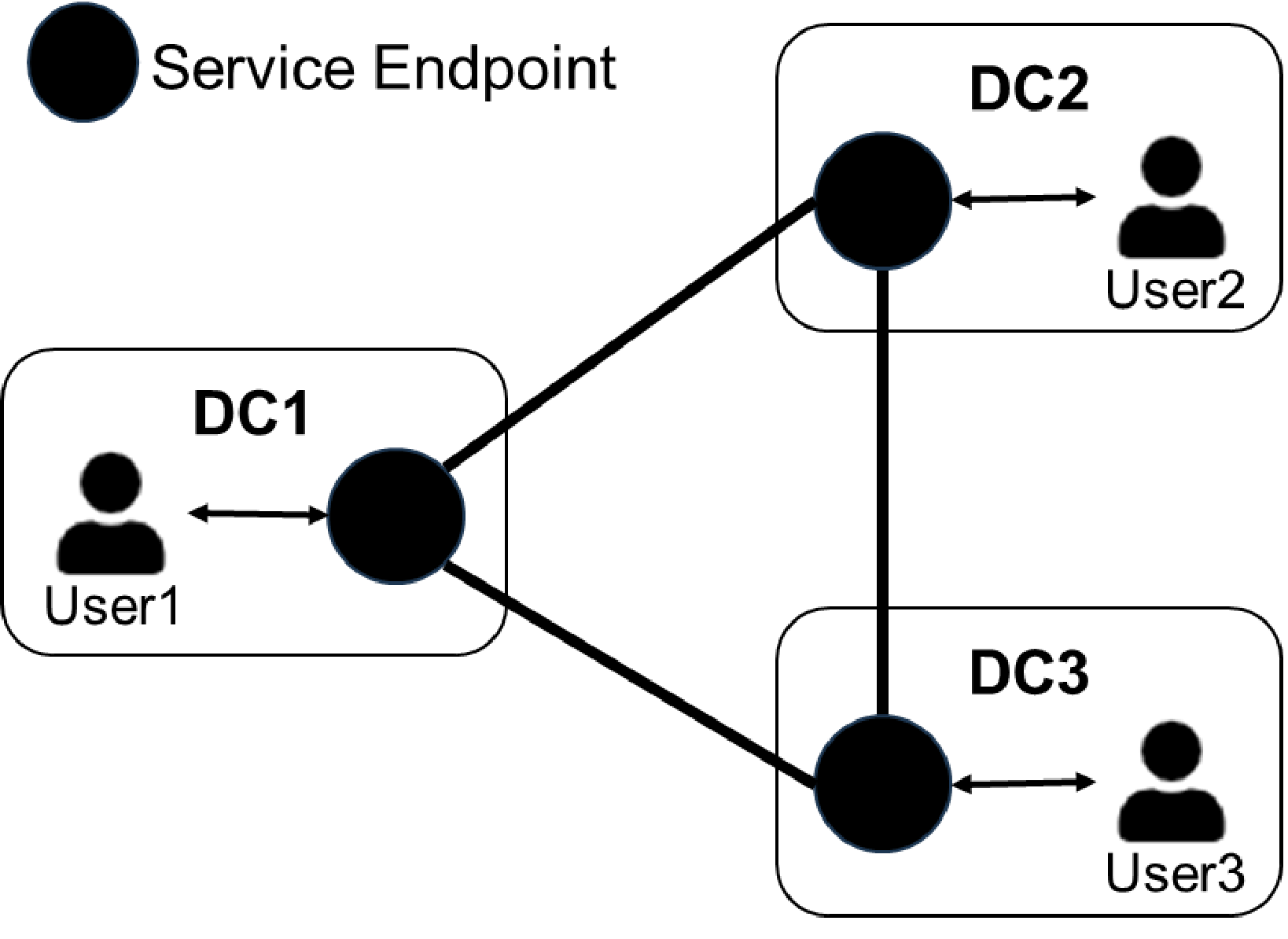}\caption{Distributed service endpoints}\label{fig:archit:b}
	\end{subfigure}\vspace{-1ex}
    \caption{video conferencing system service architectures.}
    \label{fig:archit}
    \vspace{-4ex}
\end{figure}

Due to the global distribution of users, most video conferencing systems are also architected and deployed globally to achieve good scalability and low latency. For example, Zoom has 19 interconnected data centers (DCs) spread across the world and can also scale to public clouds when needed~\cite{zoomdcs}. 
Existing video conferencing systems may have different architectures.
For example, Figure~\ref{fig:archit} shows two service architectures adopted by popular video conferencing systems~\cite{10.1145/3487552.3487847,10.1145/3487552.3487842}. 
For both architectures, large amounts of data are transferred across geo-distributed data centers, either between endpoints and users (Figure~\ref{fig:archit:a}) or between multiple endpoints (Figure~\ref{fig:archit:b}).
The existing studies show that the inter-DC network latency can be up to hundreds milliseconds (see Section~\ref{sec:bg:net}). The high variance of inter-DC network performance also cause high jitter, which leads to challenges when reducing the end-to-end latency of geo-distributed video conferencing. Below we summarize the challenges faced by existing video conferencing systems.

First, large cloud providers, with geographically dispersed data centers connected by a managed backbone network, allow the creation of efficient overlay alternatives to the default Internet path~\cite{via-sigcomm16,Xu2012,Haq2015}.
Existing studies have proposed various online~\cite{via-sigcomm16,7164897} algorithms to find the best routing for video conferencing applications. However, when making the routing decisions, existing methods mainly focus on minimizing the packet transmitting latency from source to target, while in video conferencing applications, the \emph{end-to-end latency} of a packet contains not only the \emph{data transmitting time} but also \emph{packet reordering time} sensitive to network jitter (i.e., Figure~\ref{fig:webRTCJB}). 
Due to the dynamic and heterogeneous network performance in geo-distributed DCs, it is highly possible that the routing decisions which lead to the \emph{minimum average packet transmitting latency do not guarantee the lowest average end-to-end latency}. 
To provide low end-to-end latency video conferencing, we need to redesign the packet routing approach to find routing path with low and stable transmitting latency.

Second, inter-DC network latency varies greatly overtime. This is especially true on public Internet paths~\cite{10.1145/3038912.3052560} with the fastly increasing video conferencing traffic~\cite{vctraffic}. \emph{Higher network dynamics lead to higher jitter}, hence causing more packets to arrive at the destination \textit{out-of-order}. Packets that are arriving too late will be discarded which results in packet loss to data streams. Existing systems mostly adopt jitter buffers~\cite{10.1145/2619239.2626296,webrtc} to handle the out-of-order problem, where the buffer size is an important parameter to balance between packet reordering latency and loss rate. By analyzing the jitter management in existing systems, we found that the buffer-based methods are essentially in-order-processing (IOP) methods and can cause unnecessary delay for packets in the buffer when there are stragglers. Due to the highly dynamic feature of geo-distributed networks, the IOP-based jitter buffer methods will inevitably cause more latency waste.

In this paper, we tackle two critical challenges in improving the quality of video conferencing in geo-distributed DCs by proposing two novel techniques: \emph{VCRoute} and \emph{WMJitter}.

First, VCRoute is a packet routing method specifically designed for video conferencing systems. VCRoute treats the routing problem as a Multi-Armed Bandit problem and applies a custom version of Thompson Sampling to solve it. In contrast to the previous approaches~\cite{via-sigcomm16}, 
our method leverages the Bayesian framework inherent in Thompson Sampling to derive probability distributions for individual arms. This innovative approach naturally considers transmitting latency and its variance simultaneously, which allows us to tackle the highly dynamic nature of geo-distributed network performance effectively.

 

Second, we propose WMJitter, a watermark-based Out-Of-Order (OOP) processing mechanism to handle network jitter. WMJitter leverages the advantages of using watermarks in streaming processing systems, enabling a further reduction in end-to-end delay. Leveraging window-based statistic method for real-time network jitter estimation used in WebRTC NetEQ module~\cite{webrtc}, we facilitate timely and accurate adjustments to the watermark. This efficient buffering management provided by WMJitter can further reduce end-to-end latency, thus enhancing the overall system performance by striking a balance between latency and loss rate. These two methods, VCRoute and WMJitter, offer a comprehensive and scalable solution to address the challenges of performance optimization for video conferencing systems in geo-distributed environments.

We evaluate our techniques on two sets of real-world geo-distributed environments with complementary features. By comparing to state-of-the-art routing and jitter managing approaches, we found that 1) VCRoute greatly outperforms existing approaches when the inter-DC network performance is highly heterogeneous (Section~\ref{sec:eval:tencent}); 2) watermark-based jitter management has shown great improvement on average end-to-end latency when the inter-DC network performance is highly dynamic (Section~\ref{sec:eval:wonder}). 
In summary, we make the following contributions:
\begin{itemize}[leftmargin=*]
    \item {To the best of our knowledge, we are the first to jointly consider transmitting latency and its variance during packet routing, which is important to reduce the end-to-end latency for data packets in video conferencing systems.}
    \item Existing video conferencing systems adopt buffer-based jitter handling, which may lead to unnecessary packet delay. We are the first to introduce watermark-based Out-of-Order-Processing (OOP) into the jitter management of video conferencing systems, which further reduced the end-to-end latency for data packets.  
    \item We extensively evaluated our design on two sets of geo-distributed environments with complementary features. Evaluation results have shown that our design can effectively reduce the end-to-end latency of packets by up to 44\% compared to the state-of-the-art. We believe the results are insightful for video conferencing service providers from different businesses (clouds, ISPs, etc.).
\end{itemize}

The remainder of this paper is organized as follows. We introduce the background and motivation in Section~\ref{sec:bg}, present the system overview in Section~\ref{sec:overview}, unfold design details in Section~\ref{sec:vcroute} and Section~\ref{sec:wm}, introduce our evaluation in Section~\ref{sec:eval}, present the related literature in Section~\ref{sec:related} and finally conclude the paper in Section~\ref{sec:conclude}.

\vspace{-1ex}
\section{Background and Motivation}\label{sec:bg}

\subsection{Geo-Distributed Network Features}\label{sec:bg:net}
Geo-distributed data communications go through the Wide Area Network (WAN), which is much different from intra-DC network due to its {high latency} and {high dynamicity}~\cite{7604140}. In the following, we present observations from existing studies as well as our measurement of real inter-DC network performance on Tencent cloud, to mimic the deploying environment of a geo-distributed video conferencing system using public network connections.


\begin{figure}[t]
    \centering
    \includegraphics[width=0.9\linewidth]{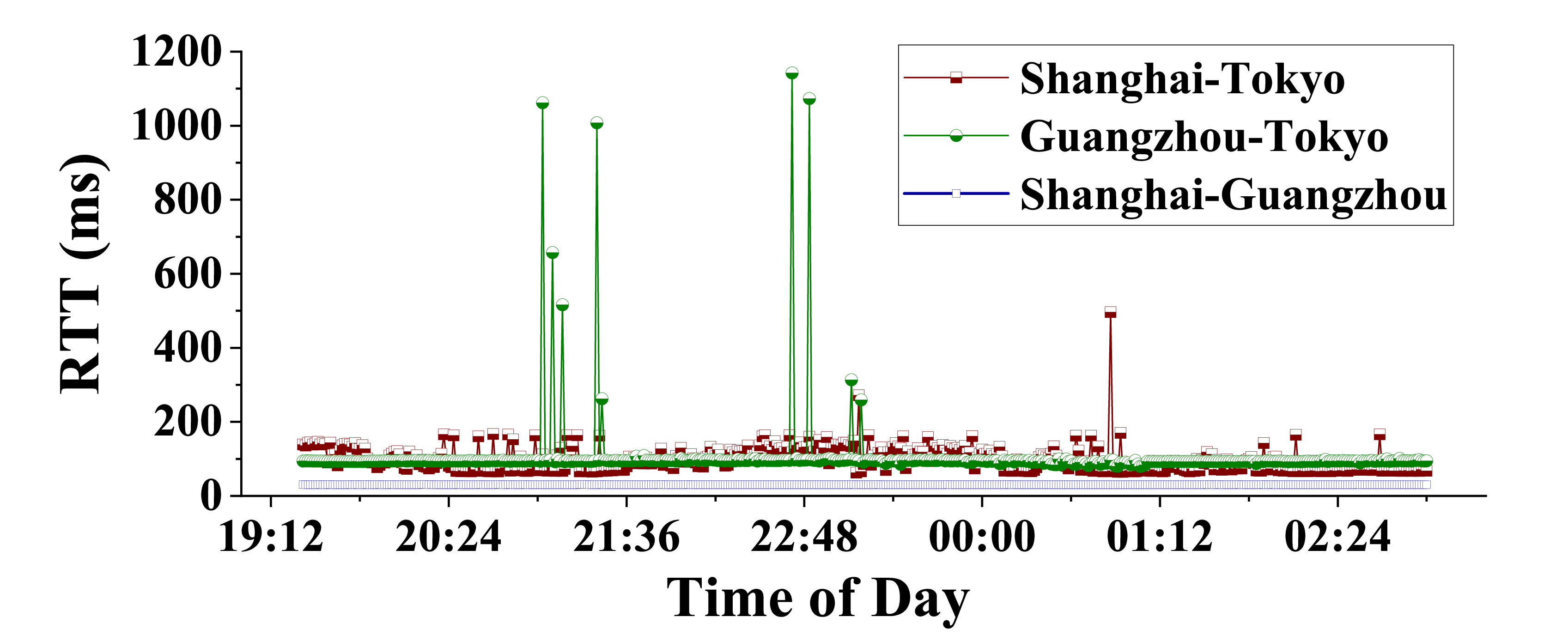}\vspace{-1ex}
    \caption{RTT variation over time between different Tencent cloud DCs. We measure RTT once per minute and plot the average of every ten minutes. }\label{fig:latency:time}
    \vspace{-4ex}
\end{figure}

First, the inter-DC network latency of different regions vary a lot. Existing measurement results of popular public clouds~\cite{AWSGCPcloud,AWScloud,tencentcloud} show that the maximum medium value of iter-DC network latency (330ms) can be up to 33 times of the minimum value. And a rich of studies~\cite{via-sigcomm16,vskyconf,Xu2012,Haq2015} have demonstrate that overlay routing is an applicable and effective way to reduce the iter-DC network latency.
Second, inter-DC network latency can vary greatly over time. We measure the RTT between two instances located in different Tencent cloud DCs for a few hours and plot the variation of RTT along time in Figure~\ref{fig:latency:time}. The highest RTT between Guangzhou and Tokyo can be over ten times larger than the average. There are also inter-DC links with low performance variations, such as that between Shanghai and Guangzhou.
This may depend on the distance between two DCs as well as the traffic on the link.

The dynamicity of network latency in the geo-distributed network can lead to \emph{jitter} problem, which refers to the variation in the delay of packets as they traverse a network. Jitter can be caused by a variety of factors, including network congestion, packet queuing, and routing issues. High levels of jitter can result in poor network performance and may cause problems for real-time applications such as voice and video conferencing.
{Although packet loss rate between inter-DC network paths can also impact the quality of real-time applications, in this paper, we assume existing techniques such as packet loss concealment~\cite{8170224} can adequately address the issue and focus mainly on the high heterogneity and dynamicity issues of network latency.}

\vspace{-1ex}
\subsection{video Conferencing Systems}

In the past few years, especially since the outbreak of COVID-19 pandemic, video conferencing systems have played an important role in our day-to-day communications. Optimizing the quality of video/audio calls is urgent considering the large number of end users distributed globally.
{Existing studies have revealed the relationship between call quality and network performance~\cite{via-sigcomm16}, which shows that network metrics such as RTT and jitter can be used to accurately measure the quality of video/audio calls. That is, any improvement in RTT or jitter is likely to improve the call quality~\cite{via-sigcomm16}. 
However, the special network features in geo-distributed DCs (e.g., high RTT, heterogeneous relay paths and high jitter) make it especially challenging to obtain good optimizations. 
}

{Popular video conferencing systems such as Zoom and Webex adopt the single service endpoint architecture shown in Figure~\ref{fig:archit:a}~\cite{10.1145/3487552.3487847}, where a single service endpoint is scheduled for each meeting session and all participants communicate with the endpoint for audio streaming.}
The endpoint is used to select and forward video/audio streams. Due to the high latency of inter-DC network, this can clearly lead to large delay for users far from the endpoint if without any routing optimizations. Whats 'more, data packets in the same stream using default Internet paths (e.g., BGP-derived) are routed individually, they may take different underlay paths and encounter varying network latency, leading to high jitter. 
As discussed in the last subsection, using other service nodes as relays and providing efficient routing policy between endpoint and users may be able to reduce the delay and jitter for packets. 

\begin{figure}[t]
    \centering
    \includegraphics[width=1\linewidth]{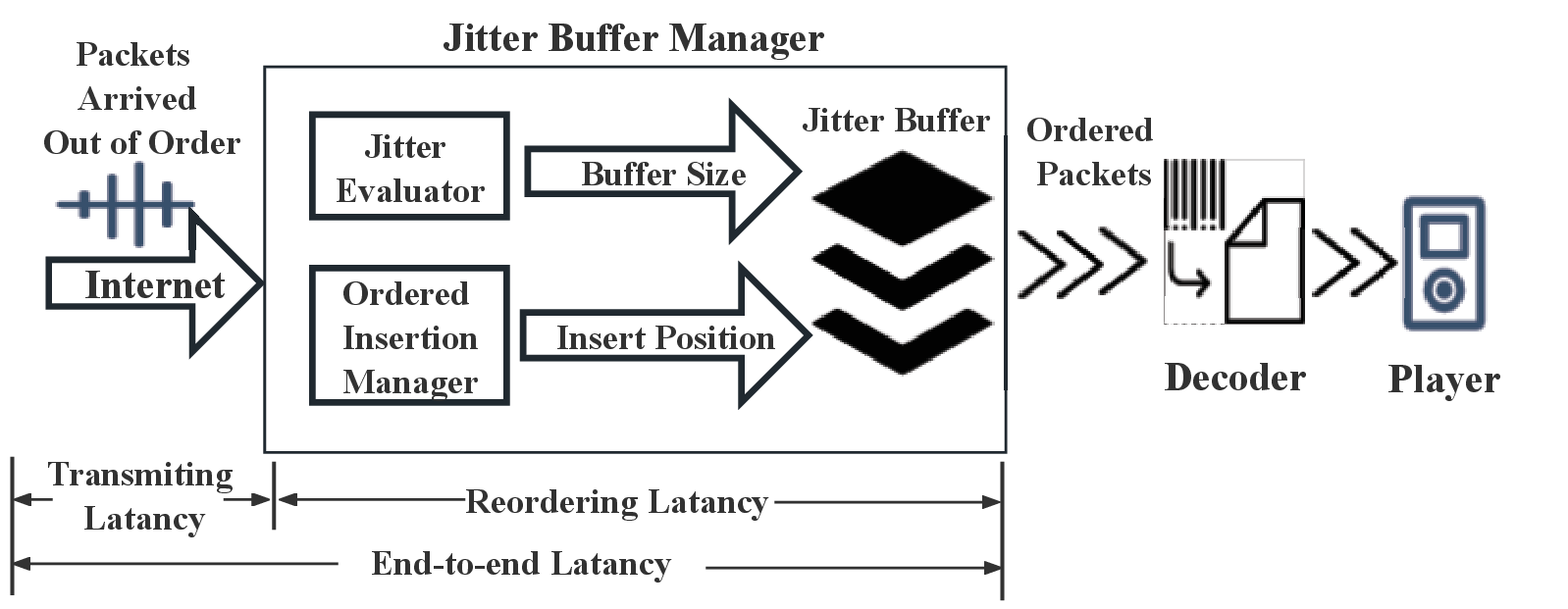}\vspace{-1ex}
    \caption{Jitter Buffer Manager in webRTC}\label{fig:webRTCJB}
    \vspace{-2ex}
\end{figure}

Existing network communication systems usually adopt {jitter buffers} to compensate the quality degradation caused by jitter. 
For example, Figure~\ref{fig:webRTCJB} shows the jitter buffer management in webRTC~\cite{webrtc}, a popular open-source framework for creating video conferencing systems.
Data packets travel through the Internet and arrive at users' end.
A jitter buffer at the user's end temporarily stores incoming packets which are usually out-of-order, and then plays them back by reordering the packets according to their timestamps of generation. Packets arriving too late are discarded. Jitter buffer absorbs the variability in packet arrival times, allowing the communication system to maintain a consistent quality of service. The buffer size is an important parameter. A large buffer can provide better jitter compensation, at the expense of increased delay, while a small buffer may cause more packets being discarded due to out-of-order. Existing studies~\cite{10.1145/2740070.2626296,6967689} proposed to adapt the buffer size according to network jitter.

\vspace{-1ex}
\subsection{Motivations}


With the global pandemic of COVID-19, some ISPs have reported a more than 300\% increase on video conferencing traffic from February to October 2020~\cite{vctraffic}.
As a result, the network congestion and variation will cause more severe impact on the performance of video conferencing systems. 

Existing studies optimize the performance of video conferencing systems by reducing the end-to-end latency of data streams.
As shown in Figure~\ref{fig:webRTCJB}, the end-to-end latency of a data stream is composed of two parts, including the time for transmitting a packet through the Internet (i.e., \emph{transmitting latency}) and the lag that a packet stays in the jitter buffer (i.e., \emph{reordering latency}).
Existing studies tried to reduce the transmitting latency and reordering latency individually through finding good dissemination paths~\cite{via-sigcomm16} or designing smart buffering schemes~\cite{vskyconf}.
However, we found that optimizing the two components individually does not guarantee good overall performance for video conferencing systems.

\begin{figure}[t]
    \centering
    \begin{subfigure}{0.32\linewidth}
    \setlength{\abovecaptionskip}{-1pt}
	\hspace{-0.55cm}
 	\includegraphics[width=1.2\linewidth]{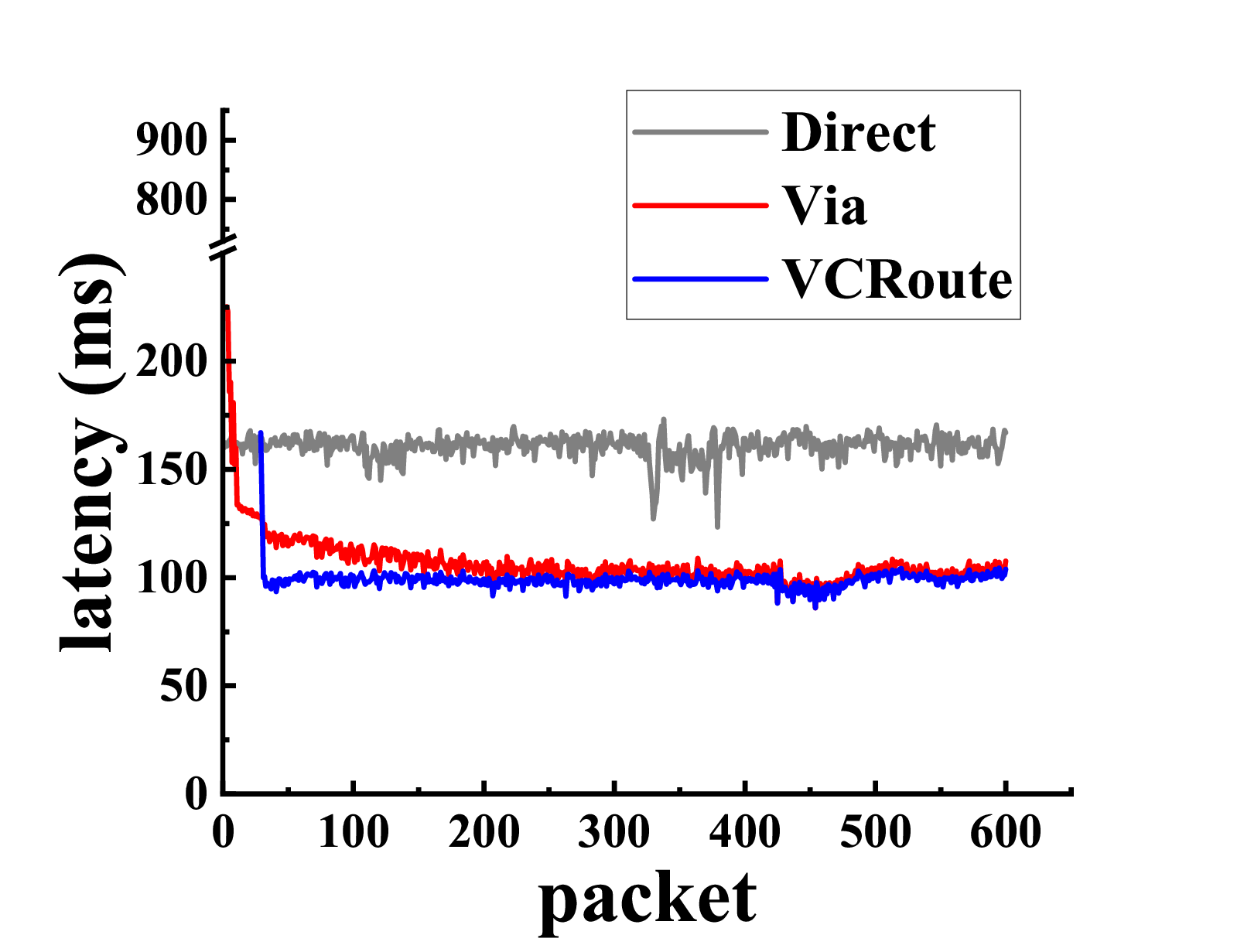}\caption{Transmitting}\label{fig:motiv1:c}
	\end{subfigure}
 \hfil
    \begin{subfigure}{0.32\linewidth}
    \setlength{\abovecaptionskip}{-1pt}
	\hspace{-0.55cm}
	\includegraphics[width=1.2\linewidth]{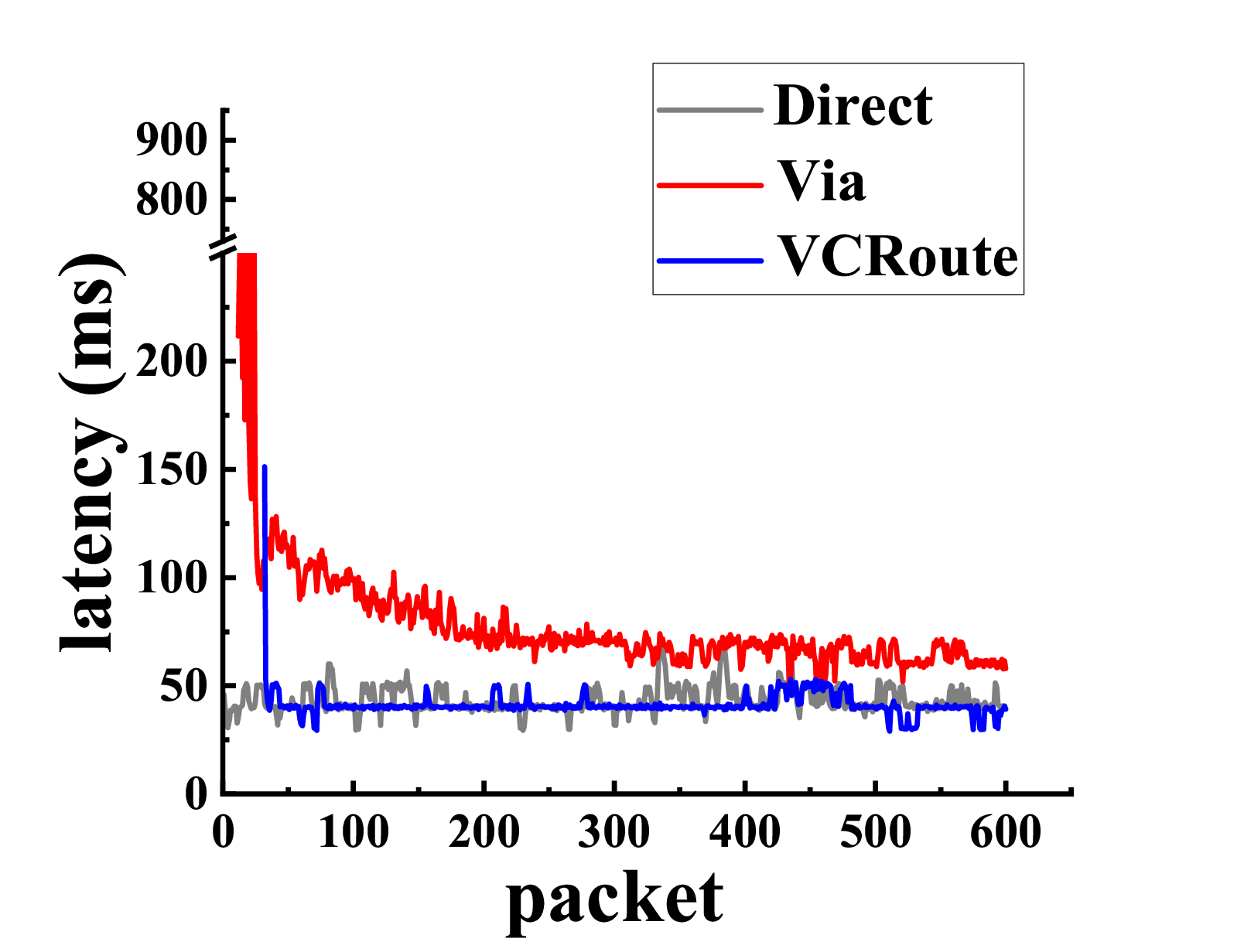}\caption{Reordering}\label{fig:motiv1:b}
	\end{subfigure}
 \hfil
    \begin{subfigure}{0.32\linewidth}
    \setlength{\abovecaptionskip}{-1pt}
	\hspace{-0.55cm}
	\includegraphics[width=1.2\linewidth]{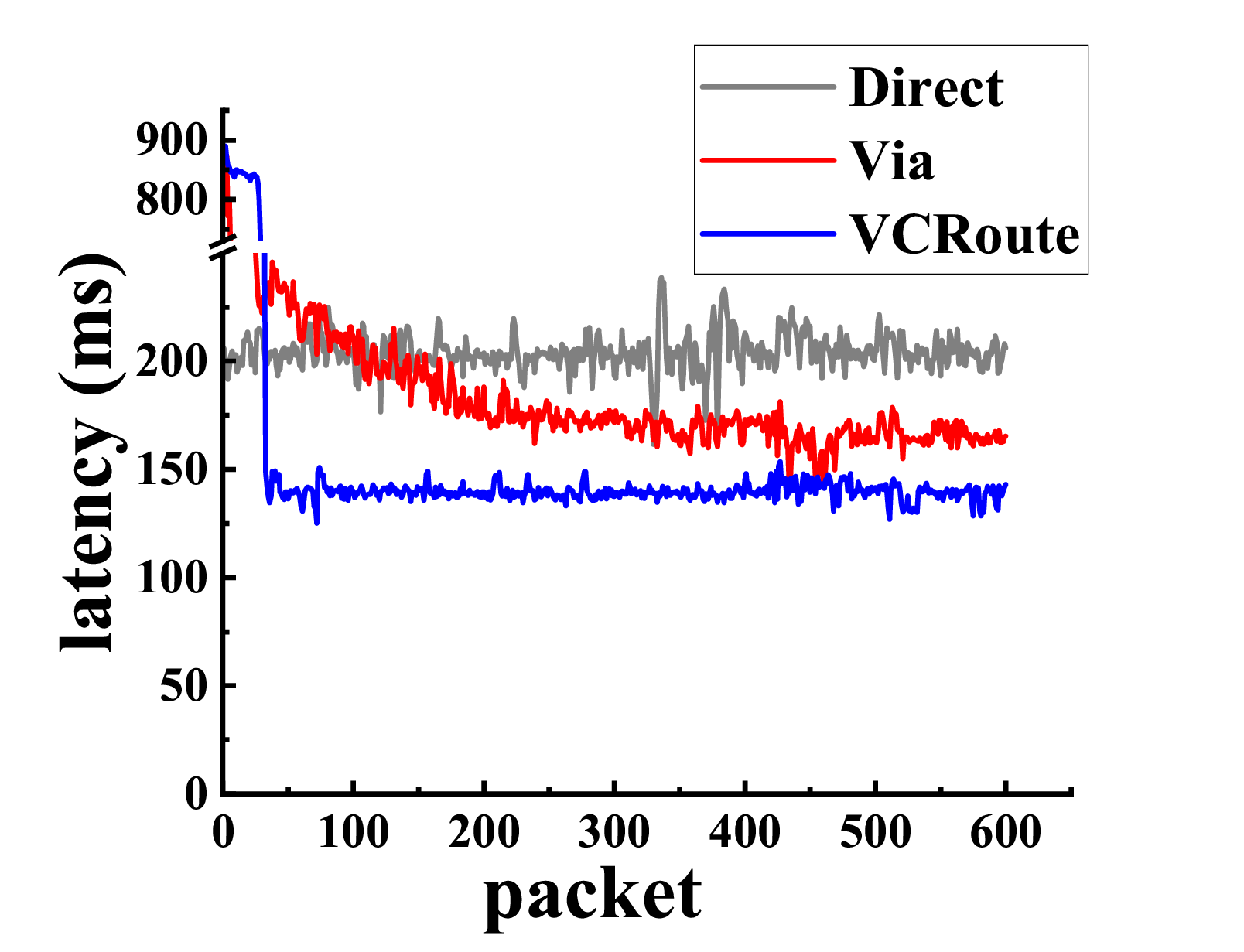}\caption{End-to-end}\label{fig:motiv1:a}
	\end{subfigure}
 \vspace{-1ex}
    \caption{Latencies of 30,000 packets communicated between two ends in a video conferencing system. Average latency of every 50 packets is plotted. 
    }\label{fig:motiv1}
    \vspace{-4ex}
\end{figure}

In a global video conferencing session, data streams are transmitted through WAN composed of multiple autonomous systems. Data packets in a stream are routed individually and thus may end up with different underlay paths. 
Existing routing methods such as Via~\cite{via-sigcomm16} mainly focus on selecting the best relay paths for packets in geo-distributed overlay networks~\cite{9546478} to minimize packet transmitting time.
However, this does not guarantee low {end-to-end latency} for packets, due to ignorance of {packet reordering lag}. 

Consider a simple example with two communicating ends and four relay nodes (17 paths between the two ends considering only direct paths and indirect paths with one or two relays). 
The network performance between each pair of nodes are adopted from real measurements in Tencent cloud. Consider a stream of 30,000 packets sent from one end to the other (emulating a 5min call). We adopt three different ways to make routing decisions for each packet in the stream, including Direct, Via~\cite{via-sigcomm16} and VCRoute. 
The same packet reordering approach is used for all comparisons. 
As shown in Figure~\ref{fig:motiv1:c}, both Via and VCRoute can find relay paths leading to much lower data transmitting time than directly communicating between the two ends (i.e., Direct). However, due to ignorance of the packet reordering lag, Via tend to explore different routes to find paths with lower transmitting latency, thus leading to higher jitter. As a result, Via has the highest packet reordering latency among the three as shown in Figure~\ref{fig:motiv1:b}. By jointly considering packet transmitting and reordering, VCRoute obtains the lowest end-to-end latency.

The above results have shown the necessity to \textbf{jointly consider network latency and jitter during packet routing in video conferencing systems}. This is especially important for geo-distributed video conferencing, where the time-variation and heterogeneity of WAN performance gets more significant.


Data packets transmitted through the WAN are usually out-of-order and need to be reordered for good user experience. Existing studies mostly adopt jitter buffer to reorder packets at the receivers' end~\cite{salsify}. 
Although one can dynamically change buffer size to adapt system latency to different network variations, jitter buffer is essentially an in-order-processing (IOP) method. The buffering mechanism inevitably causes latency penalty for all packets as they wait for stragglers to arrive~\cite{watermark-vldb}. However, the processing stage that comes after packet reordering (e.g., decoding) does not necessarily require all packets to be strictly in-order. 

In this paper, we found that it is possible to \textbf{leverage out-of-order processing (OOP) paradigm to further reduce the latency of video conferencing systems}. OOP is an architecture commonly used in streaming processing for flexible and efficient execution of stream queries~\cite{oop-vldb}. Compared to conventional jitter buffer based approaches, OOP allows more packets to come out of reordering in time while ensuring the required order, resulting in reduced latency.

In summary, special features of geo-distributed network require revisiting the routing and jitter handling in current video conferencing systems to optimize end-to-end latency.

\vspace{-1ex}
\section{System Overview}\label{sec:overview}

\begin{figure}[t]
    \centering
    \includegraphics[width=\linewidth,height=0.45\linewidth]{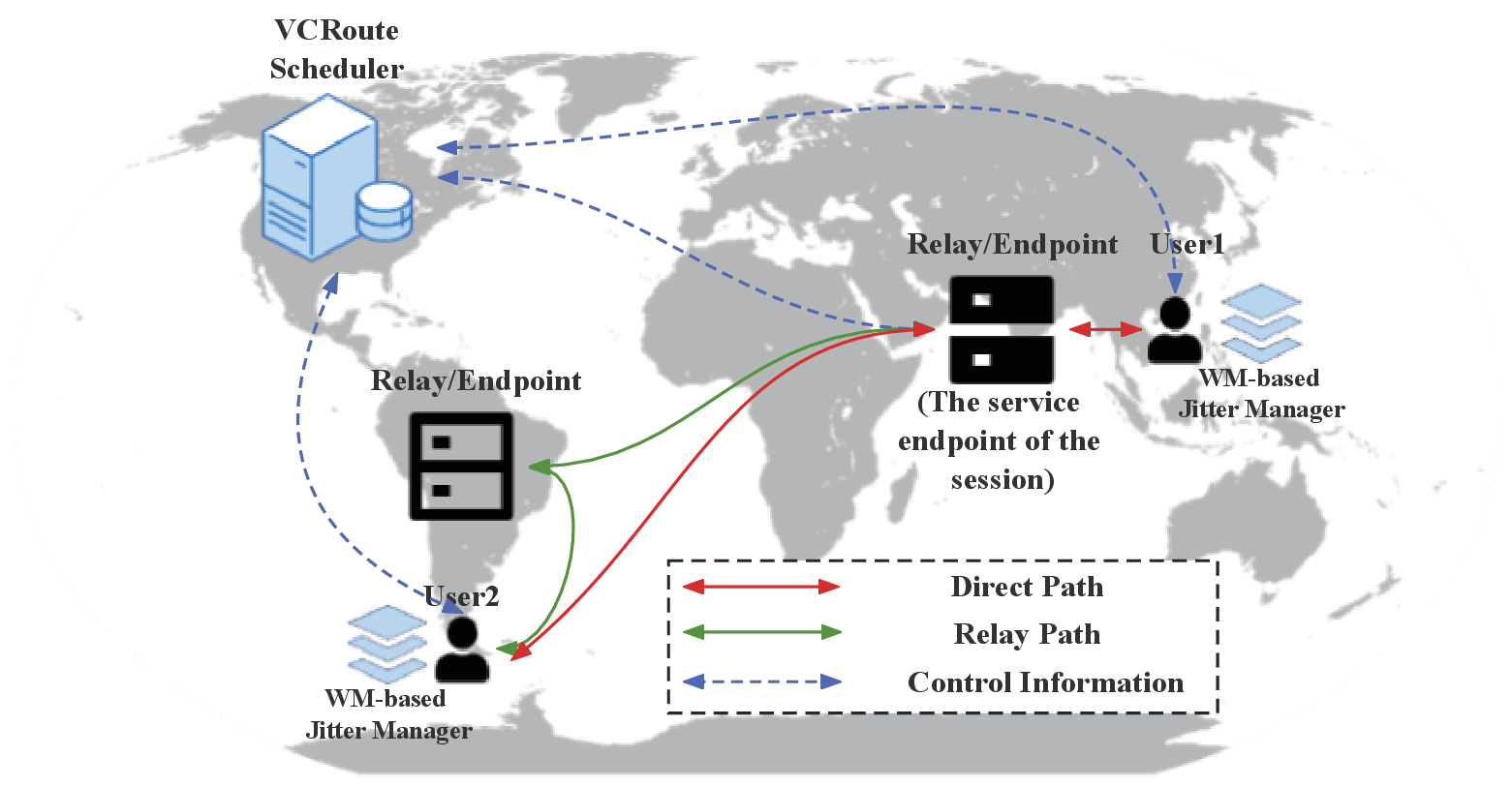}    
    \vspace{-1ex}
    \caption{System Overview}
    \label{fig:system}
    \vspace{-4ex}
\end{figure}

Motivated by the deficiencies of existing work, this paper aims at proposing a \emph{low-latency} video conferencing system for geo-distributed DCs. We make two efforts to achieve this goal. First, we designed an {application-aware} routing mechanism named \emph{VCRoute}, which jointly considers the network latency and jitter during routing to reduce the end-to-end latency for data packets. 
Second, we designed a novel {jitter manager} named \emph{WMJitter} at the users' end, which utilizes the Watermark-based OOP mechanism to make sure packets come out of the jitter manager in the correct order under low latency. Figure~\ref{fig:system} gives an overview of our system. 


\textbf{Application-aware packet routing.}
Existing routing studies mainly focus on reducing the time that a packet travels on the route, and thus do not guarantee low {end-to-end} latency for video conferencing applications.
In this paper, we propose {VCRoute}, which is a centralized packet scheduler specifically designed for optimizing the performance of video conferencing systems.
As shown in Figure~\ref{fig:system}, VCRoute aims at selecting the best relay path for each packet travelling between the endpoint and the end user.


We formulate the routing problem as a Multi-Armed Bandit (MAB) problem, where each available path in the geo-distributed environment is an arm of the bandit.
Traditional MAB algorithms used in existing studies~\cite{via-sigcomm16} selects average transmitting latency as the sole optimization objective for routing. However, the high dynamicity of inter-DC network latency leads to high variance in the rewards, which can significantly impact the performance and effectiveness of these algorithms.
To address the above challenges, we carefully adapt the Thompson Sampling algorithm~\cite{thompson} to our MAB problem. Thompson Sampling can estimate the distribution function of transmitting latency for each arm, allowing for the simultaneous consideration of both transmitting latency and its variance when routing, which can effectively handle uncertainty and high variance in the rewards of arms.

\textbf{Low-latency jitter management.}
Network jitter determines the degree of disorder of the packets. The purpose of jitter management is to absorb the impact of jitter as much as possible, under a low latency.
As mentioned above, existing buffer-based jitter management is essentially doing in-order processing and thus can cause unnecessary delay of packets in the buffer.
In this paper, we propose watermark-based jitter management~\cite{watermark-vldb} at users' end. Watermarks are quantitative markers associated with a stream, indicating that no events in the stream will have a timestamp preceding that of the watermark.
Thus, we can use watermarks to safely determine when to emit packets and keep track of the progress of packet handling. No extra latency will be introduced due to existence of straggler packets.

\vspace{-1ex}
\section{VCRoute Scheduler}\label{sec:vcroute}
VCRoute provides a centralized routing scheduler for each packet in the audio streams, so that the end-to-end latency of the packets can be minimized.
To achieve this goal, VCRoute needs to answer the following questions:
\begin{itemize}[leftmargin=*]
    \item First, as VCRoute is designed to be centralized, how can we mitigate the system overhead introduced by transmitting scheduling decisions to users and endpoints? (Section~\ref{sec:vcroute:overhead})
    \item Second, given the high latency and dynamicity of inter-DC network performance, how can VCRoute find a good path for each packet in a timely manner? (Section~\ref{sec:vcroute:pred})
\end{itemize}
We have carefully designed VCRoute to solve the above questions. Figure~\ref{fig:routing:overview} gives an overview of VCRoute.

\vspace{-1ex}
\subsection{Reducing overhead}\label{sec:vcroute:overhead}
Many existing routing algorithms such as Via make one routing plan for an entire call at offline time. Thus, the scheduling overhead is not a major concern.
However, this is not good enough for our problem due to the highly dynamic and heterogeneous network performance in geo-distributed video conferencing.
For example, in Figure~\ref{fig:motiv1}, the Direct method which uses a fixed routing plan for an entire stream leads to high packet transmitting time. 
On the other hand, making different routing plans for each individual packet leads to high scheduling overhead. We address the overhead issue from two aspects. 

\begin{figure}[t]
    \centering
    \includegraphics[width=0.9\linewidth]{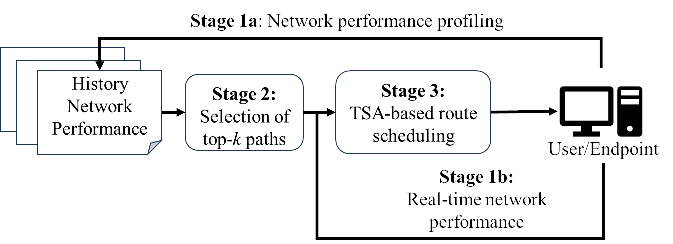}
    \caption{Overview of VCRoute}\label{fig:routing:overview}
    \vspace{-3ex}
\end{figure}

First, for each communicating $<$endpoint, user$>$ pair, we set a default routing plan. All packets in the stream are scheduled according to the default plan. VCRoute updates the default routing plan according to real-time network performance. Thus, we only need to transmit the routing decisions when they are updated. As a result, we are able to greatly reduce the overhead of the centralized scheduler.
For example, in Figure~\ref{fig:motiv1}, VCRoute transmits only 447 times of routing plans for 30,000 packets, resulting in a 1.5\% extra overhead compared to a decentralized design.
Further, we found that among the 447 times of routing plan updates, only 4 paths out of 17 were selected. This motivates our second optimization of 
selecting the top-k low-latency paths from all relay paths to reduce the solution space (Stage 2 of Figure~\ref{fig:routing:overview}). The top-k paths are obtained based on the analysis of long-term network performance history. Similar to Via~\cite{via-sigcomm16}, the k value is dynamically decided based on confidence bounds for each relay on a specific $<$endpoint, user$>$ pair (refer to~\cite{via-sigcomm16} for more details).
By using these confidence bounds to dynamically decide the top-k relaying options, it ensures that any relay option not included in the top-k is worse than any that is with a high degree of confidence.
Given the pruned set of relay paths, VCRoute selects the path that results in the best end-to-end latency at runtime, with a much lower overhead.

\vspace{-1ex}
\subsection{Making good routing decisions}\label{sec:vcroute:pred}

Given the top-k paths, we design an online routing method to dynamically select the best path for each packet (Stage 3 of Figure~\ref{fig:routing:overview}).
As mentioned previously, selecting the best path from the k options can be defined as a classic MAB problem. Each path can be considered as an arm of the bandit, and the transmitting latency of the path is used as the reward of the arm. The MAB problem has been extensively studied. However, the high-variance and dynamically changing reward distributions of the bandits could make traditional MAB algorithms such as UCB1~\cite{ucb1,via-sigcomm16} to choose path with low average transmitting latency but high variance.

\subsubsection{Thompson Sampling algorithm}
In this paper, we carefully adapt the Thompson Sampling algorithm (TSA)~\cite{thompson} to solve our MAB problem. TSA uses a Bayesian framework to model the uncertainty in reward distributions of arms. It starts from a guess that the rewards follow a prior distribution. When getting more evidence (real rewards), the TSA iteratively updates the hyperparameters of the prior and use them to maintain a posterior distribution following the same parametric form as the prior distribution. In our problem, we assume that the likelihood of the reward of each arm (i.e., transmitting latency of each path) follows a Normal distribution with known variance. The variance can be calculated using historical calls collected offline. Then, according to Bayesian probability theory, the prior and posterior probability distributions both follow the Normal distribution with hyperparameters mean $\mu$ and variance $ \sigma^2$. We use the observed rewards to update the hyperparameters for the posterior distribution.



Below we introduce how the hyperparameters are updated. 
For simplicity, we use precision $ \tau $ ($ \tau  = 1/\sigma^2 $) instead of the variance $ \sigma^2$ to update the posterior distribution.
According to Bayesian probability theory, we have:
\begin{eqnarray}
\tau \gets \tau+n*\tau_0 
\end{eqnarray}\vspace{-4ex}
\begin{eqnarray}
\mu \gets (\tau*\mu+\tau_0*  {\textstyle \sum_{i}^{n}{x_i}})/(\tau+n*\tau_0)
\end{eqnarray}
{where $ \tau_0 $ is the precision of calculated using historical end-to-end latency traces, $ n $ is the number of times the arm has been tested and $ x_i $ is the end-to-end latency received at the $ i $-th test  of the arm. 
}

At each time step, TSA samples reward from the posterior predictive normal distribution of each arm and select the arm with the highest returned value. 
TSA transparently combines both exploration and exploitation. Once an arm is tested and a reward is obtained, the belief in the likelihood of the reward of that arm is modified. 
By employing sampling strategies based on posterior probabilities associated with the optimality of different actions, the algorithm ensures continued exploration of all actions that hold plausible optimality potential while reducing sampling emphasis on actions deemed less likely to be optimal. 

\vspace{-1ex}
\section{Watermark-based Jitter Manager}\label{sec:wm}
The jitter manager sits on packet receivers' side to ensure ordered emit of packets.
To achieve low-latency audio streaming, we propose a novel jitter management method named WMJitter using the watermark-based OOP mechanism.
We mainly address the following challenges: 
\begin{itemize}[leftmargin=*]
    \item No existing studies have considered watermarks for jitter management in video conferencing systems. How can we use watermarks to handle disorder packets with low latency? (Section~\ref{sec:wmjitter})
    \item Existing buffer-based jitter management adapt buffer sizes to trade-off buffering delay and packet loss. How can we achieve the same goal using watermarks? (Section~\ref{sec:window})
\end{itemize}

\floatname{algorithm}{Algorithm} 
\renewcommand{\algorithmicrequire}{\textbf{Input:}}  
\renewcommand{\algorithmicensure}{\textbf{Output:}}  
\begin{algorithm}[t]\scriptsize
	\caption{Main process of WMJitter.}\label{alg:wm}
	\begin{algorithmic}[1]
            \State {$ Q, O\gets Null$;}
		\State {When a packet $ p $ arrives: }
            \If{$ p.ts < wm $}
                \State {Discard $ p $ and return;}
            \EndIf
            \State {$Q\cup \left \{ p \right \} $;}	
		\State {$lag = CalculateLag(p)$;}
            \State {$ wm = UpdateWM(wm,p,lag)$;}
            \For {$ p $ in $ Q $}
                \If{$ p.ts < wm $}
                    \State {$O\cup \left \{ p \right \} $ and $Q.pop(p)$;}
                \EndIf
            \EndFor
            \State {Sort and output packets in $ O $;}
	\end{algorithmic}\vspace{-1ex}
\end{algorithm}

\vspace{-2ex}
\subsection{Incorporating Watermarks in Jitter Management}\label{sec:wmjitter}

We propose the integration of watermarks to provide precise indications of streaming packet progress. Specifically, any newly arriving packet with a timestamp greater than the watermark is deemed timely, while packets arriving later are considered tardy and subsequently discarded. Timely packets are retained in the buffer, awaiting output. Upon updating the watermark, packets in the buffer that possess timestamps lower than the new watermark are sorted and output collectively. Our watermark-based OOP method effectively reduces the reordering time and waiting time compared to the conventional approach of selecting the packet with the smallest timestamp when the buffer reaches its capacity.

Algorithm~\ref{alg:wm} shows the main process of WMJitter.
Specifically, when a packet arrives at the receiver, the jitter manager first estimates the jitter of the packet transmitting lag (Line 6) and updates the watermark accordingly (Line 7).
It then determines which packets can safely be sorted and output by comparing the updated watermark and the timestamps of the packets that have not yet been output (Line 8-11).
In this way, WMJitter can guarantee ordered delivery of packets.

\vspace{-1ex}
\subsection{Balancing Latency with Packet Loss}\label{sec:window}

To dynamically strike a balance between latency and packet loss rate, we propose a novel event-driven watermark generation method (i.e., the $UpdateWM()$ function in Algorithm~\ref{alg:wm}) that adapts to the changing network conditions.

The $UpdateWM()$ function uses an adaptive watermark generation algorithm to indicate the progress of data streams affected by network jitter. Upon the arrival of a new packet, we calculate an alternative watermark by subtracting the timestamp with a designated lag, subsequently selecting the larger value between the alternative watermark and the current watermark as the updated watermark.
Formally:
\begin{eqnarray}\label{eq:wm}
WM=\max_{}{(LPWM, p.ts-lag)}
\end{eqnarray}
where $LPWM$ (last published watermark) represents the watermark before the update, $p.ts$ represents the timestamp of the new packet, and $lag$ represents the tolerated delay for late data due to network fluctuations. 
The max operation in the equation ensures that the watermark is non-decreasing, indicating that packets are progressing with time.

The $lag$ is a key parameter that directly impacts the balance between buffering delay and loss rate.
A larger lag leads to increased buffering delay and reduced loss rate, while a smaller lag leads to an opposite scenario. Intuitively, a lag value that closely approximates the real-time jitter of transmitting latency optimizes the balance between buffer delay and loss rate. To enable adaptive and timely watermark adjustment, we employ the window-based statistical method of estimating jitter used in WebRTC NetEQ module~\cite{webrtc} to determine the $lag$ value. 

\section{Evaluation}\label{sec:eval}
To evaluate the effectiveness of our design on reducing the end-to-end latency of video conferencing systems, we perform experiments using network performance traces from real geo-distributed environments.
Our trace-driven simulator is implemented based on WebRTC~\cite{webrtc}.


\subsection{Experimental Setup}

\begin{table}\footnotesize
    \centering
    \renewcommand{\arraystretch}{0.8}
    \caption{Location of endpoints and users }\vspace{-2ex}
    \begin{tabular}{M{1.2cm}M{2.5cm}M{2.5cm}}
    \toprule
         Provider& Endpoint Location & User Location\\ \midrule
         Tencent cloud& Tokyo, Guangzhou, Singapore, Frankfurt, Sao Paulo & Shenzhen, Singapore, Hong Kong\\  \midrule
         Wonder Proxy& Brazil, Australia, Singapore, South Africa, Greece & Argentina, Indonesia, Uganda, United States, Lithuania\\               
         \bottomrule
    \end{tabular}
    \label{tab:testbed}
    \vspace{-2ex}
\end{table}

\textbf{Geo-distributed setting.}
We perform our experiments based on the network performance traces collected from Tencent cloud~\cite{tencentcloud} and WonderProxy~\cite{wonderproxy}, as shown in Table~\ref{tab:testbed}.

With Tencent cloud, we select five data centers as the geo-distributed environment. One S5.Medium2 instance is launched in each data center as the endpoint/relay point. Three users are setup locally using servers located in Shenzhen, Hong Kong and Singapore (outside of Tencent cloud). 
To measure the network performance, we send one packet every 10ms between each pair of servers and record the round trip time (RTT). Each measurement lasts for 100 minutes.

WonderProxy provides a global network of 250 servers for businesses to simulate web traffic from different countries and regions. We adopt five servers as endpoints/relays and five servers as users, as shown in the table. The latency between each pair of servers are adopted from the one day measurement using Ping, open-sourced by the provider~\cite{wonderproxy}.

\begin{figure}
    \centering
      \begin{subfigure}{0.45\linewidth}
    \setlength{\abovecaptionskip}{-1pt}
	\hspace{-0.65cm}
 	\includegraphics[width=1.15\linewidth]{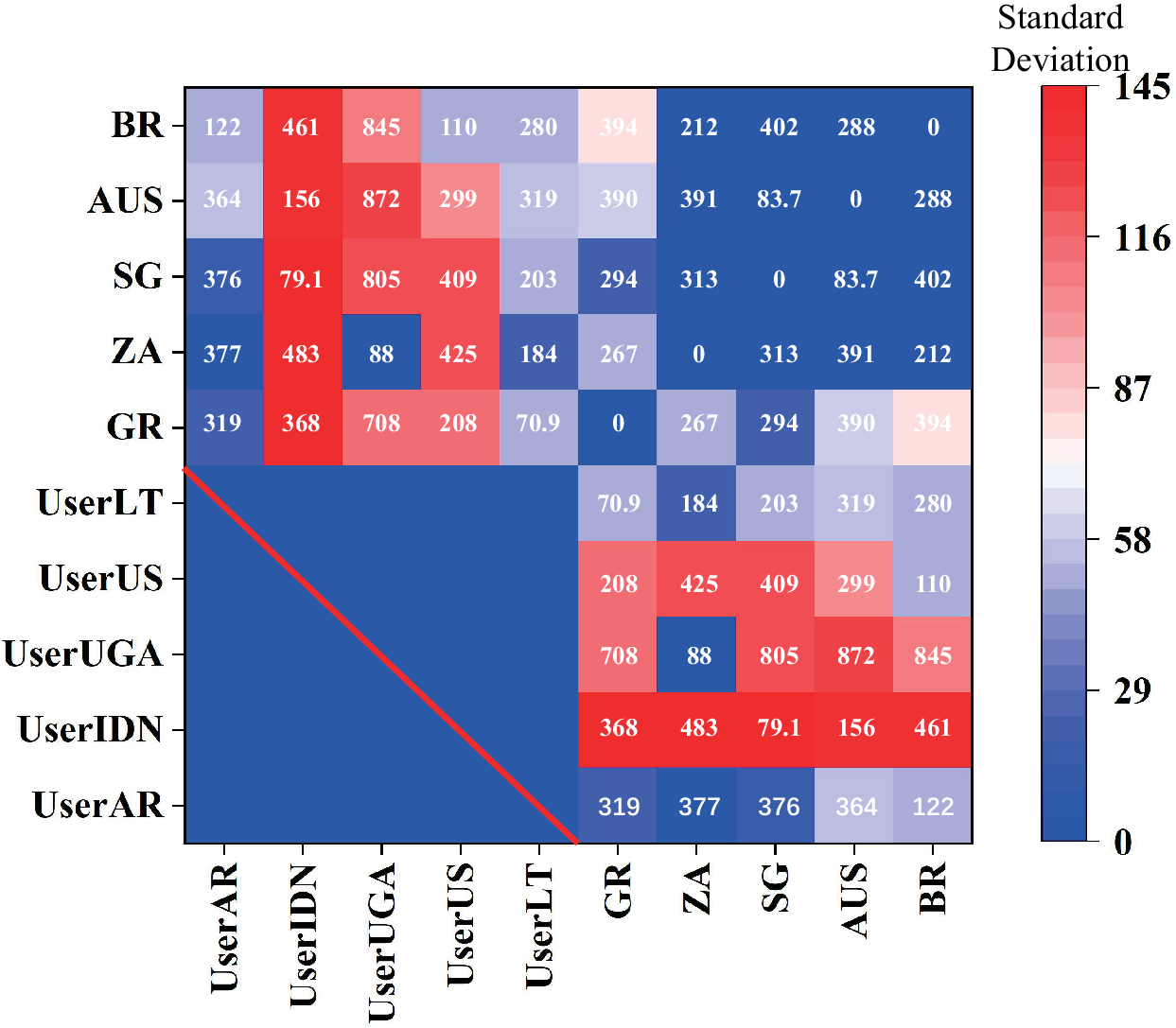}\caption{WonderProxy}\label{fig:net:wonder}
	\end{subfigure}
 \hfill
   \begin{subfigure}{0.45\linewidth}
    \setlength{\abovecaptionskip}{-1pt}
	\hspace{-0.65cm}
        \vspace{0.1cm}
 	\includegraphics[width=1.13\linewidth]{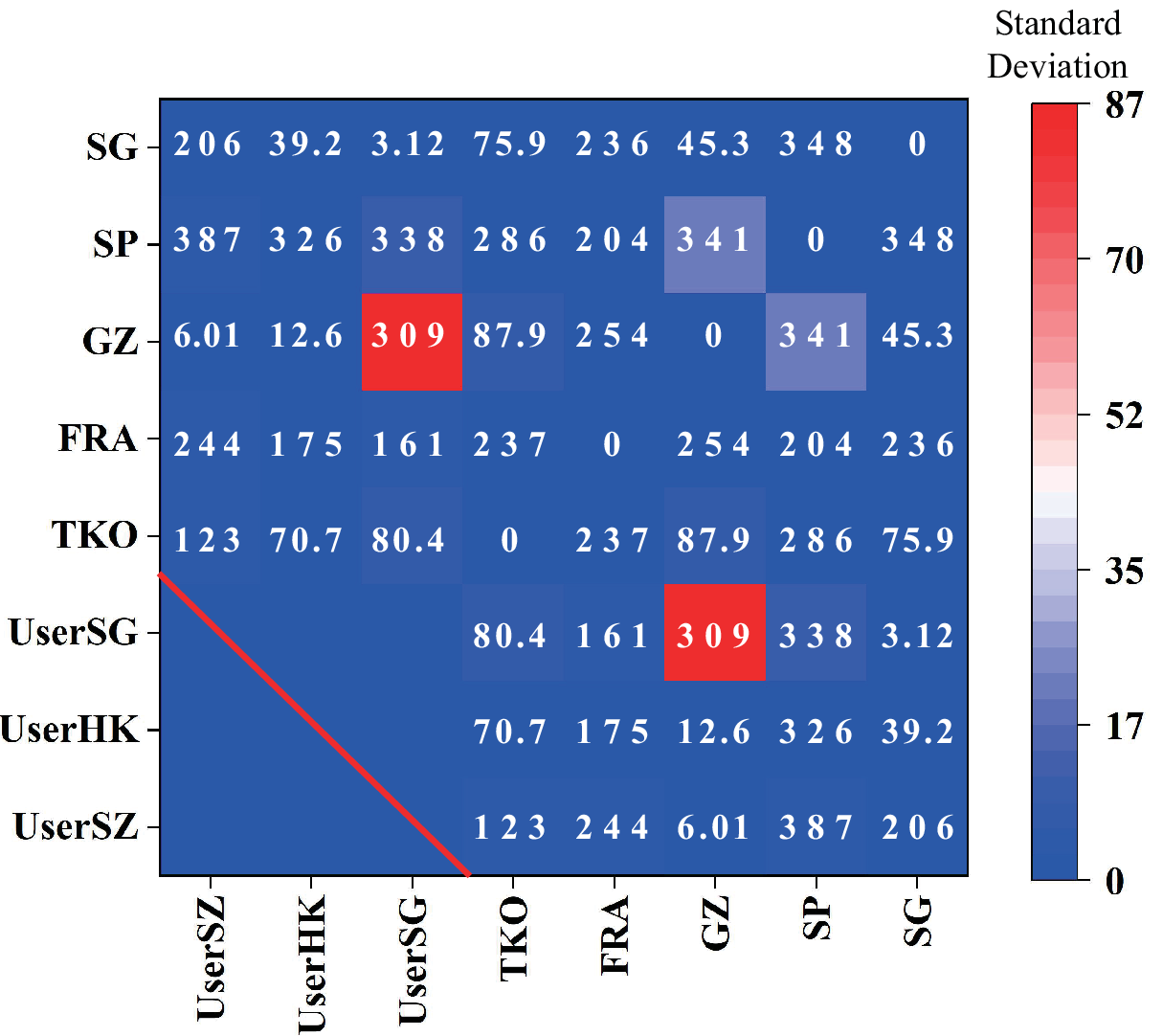}\caption{Tencent Cloud}\label{fig:net:tencent}
	\end{subfigure}
 \vspace{-1ex}
    \caption{Average network performance and variance on two platforms. Values in the figure represent average latency (ms) and the temperature represent standard deviation.}
    \label{fig:net}
 \vspace{-2ex}    
\end{figure}

As shown in Figure~\ref{fig:net}, which illustrates the average network performance and variance, the two platforms have very different features. First, the latency between WonderProxy servers are higher and more dynamic than Tencent cloud. Second, the network latency between different $<$source, destination$>$ pairs of WonderProxy are less heterogeneous. Since the two platforms differ a lot on the network features, we believe our trace-driven evaluation results are representative to show the effectiveness of our proposed algorithms.

\textbf{Compared approaches.}
To show the superiority of our methods, we adopt three routing methods and two jitter management approaches for comparison. 
{End-to-end latency is the main metric for comparing different approaches.}

We evaluate the following routing methods:
\begin{itemize}[leftmargin=*]
    \item \textbf{Direct} is the baseline routing method that transmits a packet directly from the source to target (DRT for short).
    \item \textbf{Via}~\cite{via-sigcomm16} is an online routing algorithm which selects the best overlay path for each meeting session to minimize the data transmitting latency. For fair comparison, we use it to make the best routing decision for each packet. Via is considered as the state-of-the-art comparison.
    \item \textbf{VCRoute} is the routing method proposed in this paper that minimizes the end-to-end latency for each single packet (VCR for short).
\end{itemize}

We consider the following jitter management approaches:
\begin{itemize}[leftmargin=*]
    \item \textbf{Jitter Buffer} represents the standard jitter buffer algorithm used in the NetEQ module of WebRTC~\cite{webrtc}.
    The buffer size is dynamically changed to accommodate the varying jitter conditions (BF for short).
    \item \textbf{WMJitter} is the watermark-based jitter management method proposed in this paper (WM for short). 
\end{itemize}
Putting it all together, we compare five different system solutions, namely DRT-BF, DRT-WM, VIA-BF, VIA-WM and VCR-WM. VCR-WM stands for the optimized video conferencing system of this paper.

\textbf{Configurations.}
For each geo-distributed setting, we simulate five meeting sessions hosted on the five endpoint servers. When one endpoint is selected to host a meeting, the other four endpoints are adopted as relays. As we only consider relay routes with less than two hops, there are 17 possible paths for each $<$endpoint, user$>$ pair.
For each meeting, 600,000 packets are sent between each $<$endpoint, user$>$ pair to simulate a meeting session of 100 minutes long.
During the packet handling, some late packets are dropped and cause packet loss. We report the loss rate and the end-to-end latency of successful packets as evaluation metrics.

\begin{figure*}[!htb]
    \centering
    \begin{subfigure}{0.18\linewidth}
   \setlength{\abovecaptionskip}{-1pt}
	\hspace{-0.23cm}
 	\includegraphics[width=1.2\linewidth]{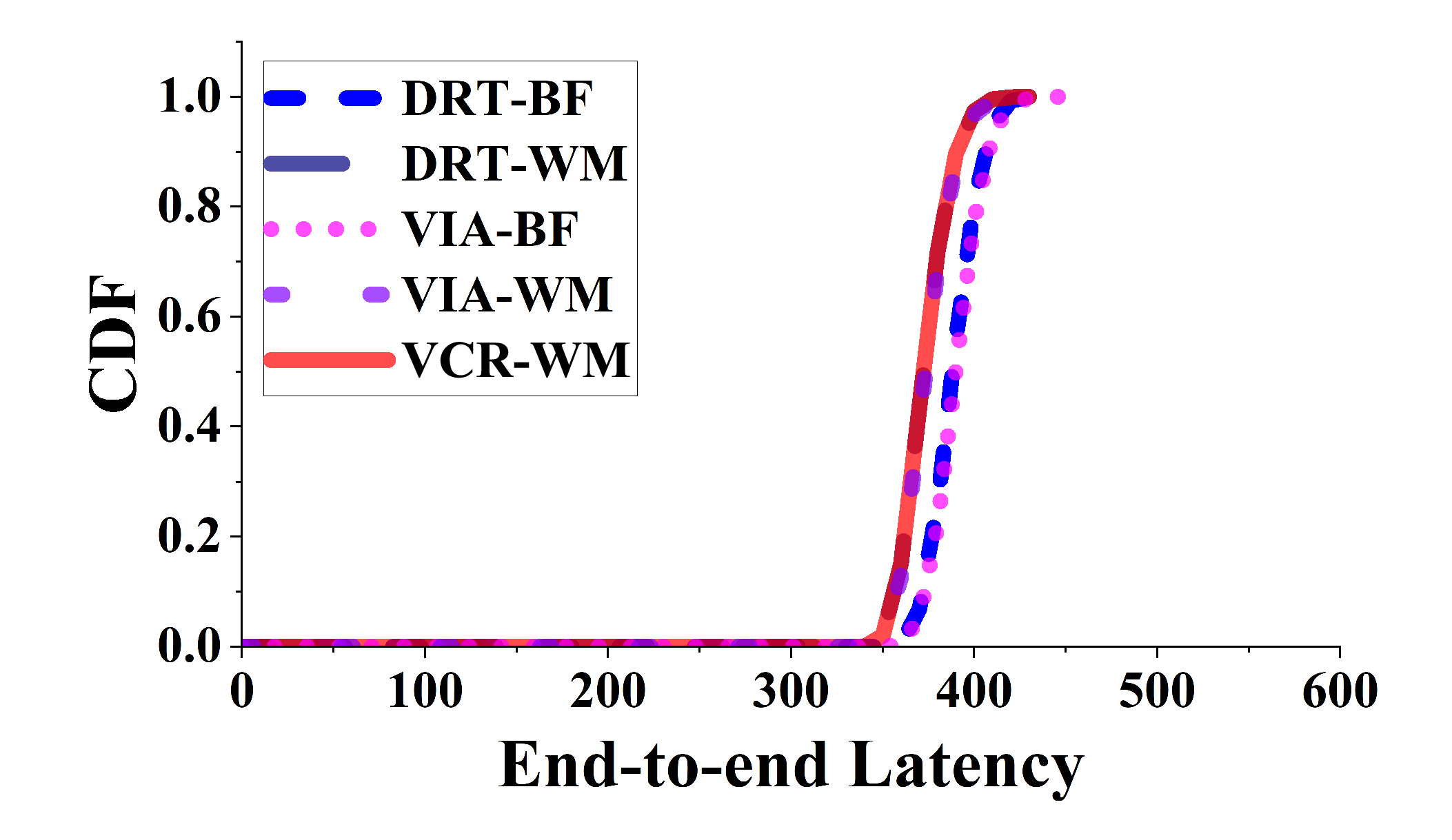}\caption{GR-UserAR}\label{fig:overall:gre_arg}
	\end{subfigure}
 \hfil
    \begin{subfigure}{0.18\linewidth}
    \setlength{\abovecaptionskip}{-1pt}
	\hspace{-0.23cm}
	\includegraphics[width=1.2\linewidth]{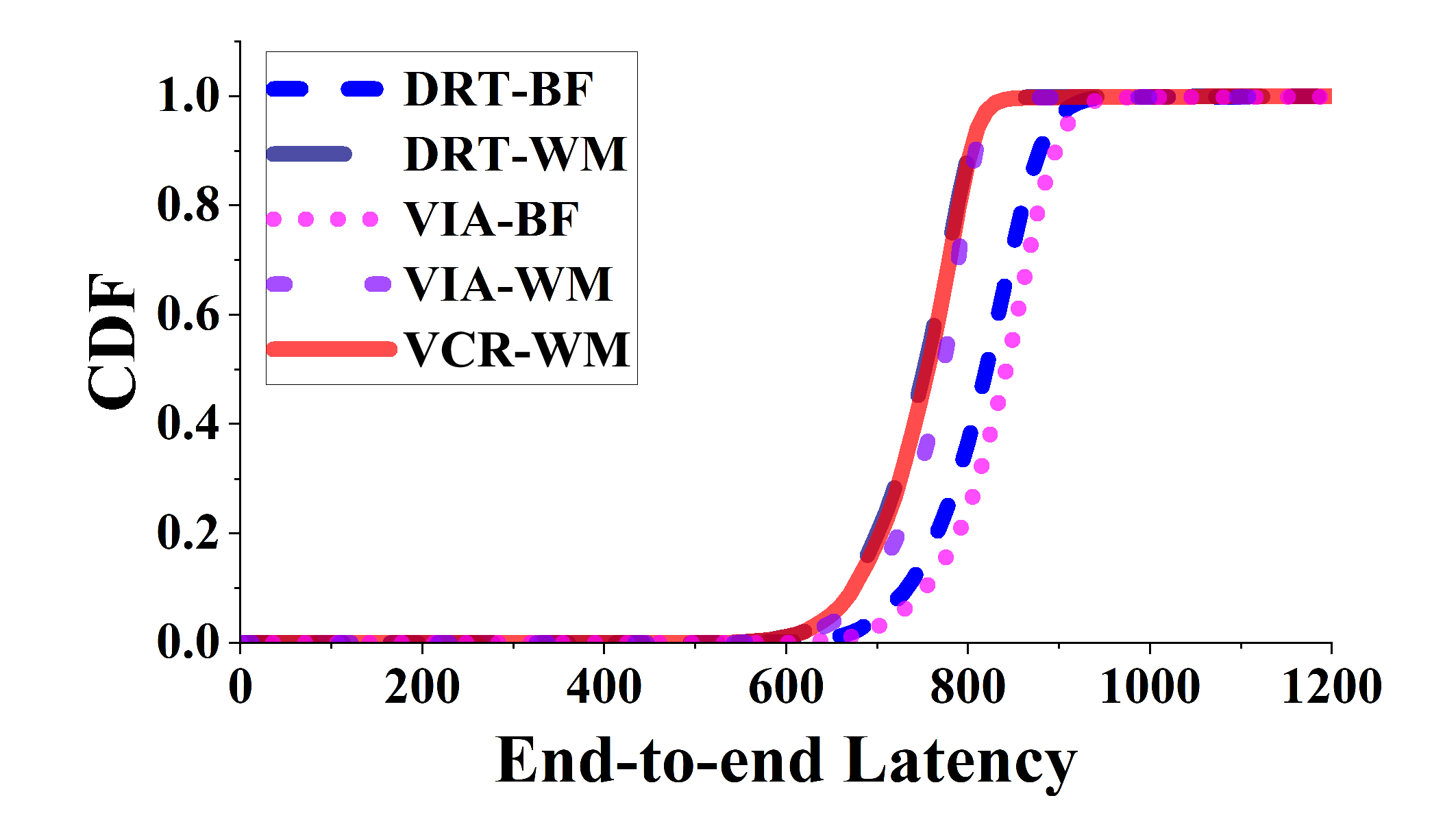}\caption{GR-UserIDN}\label{fig:overall:gre_ind}
	\end{subfigure}
 \hfil
    \begin{subfigure}{0.18\linewidth}
    \setlength{\abovecaptionskip}{-1pt}
	\hspace{-0.23cm}
	\includegraphics[width=1.2\linewidth]{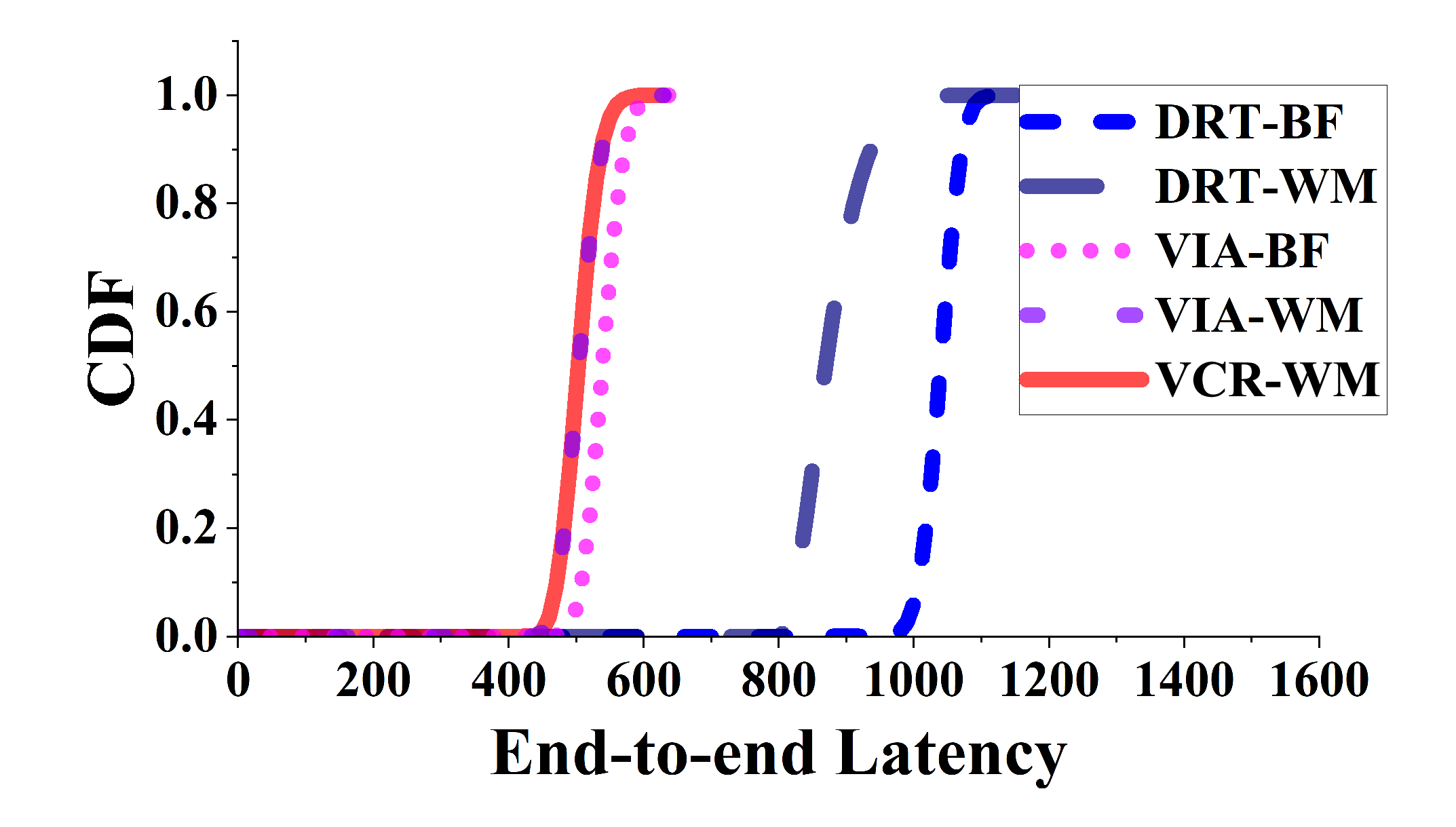}\caption{GR-UserUGA}\label{fig:overall:gre_uga}
	\end{subfigure}
 \hfil
    \begin{subfigure}{0.18\linewidth}
    \setlength{\abovecaptionskip}{-1pt}
	\hspace{-0.23cm}
	\includegraphics[width=1.2\linewidth]{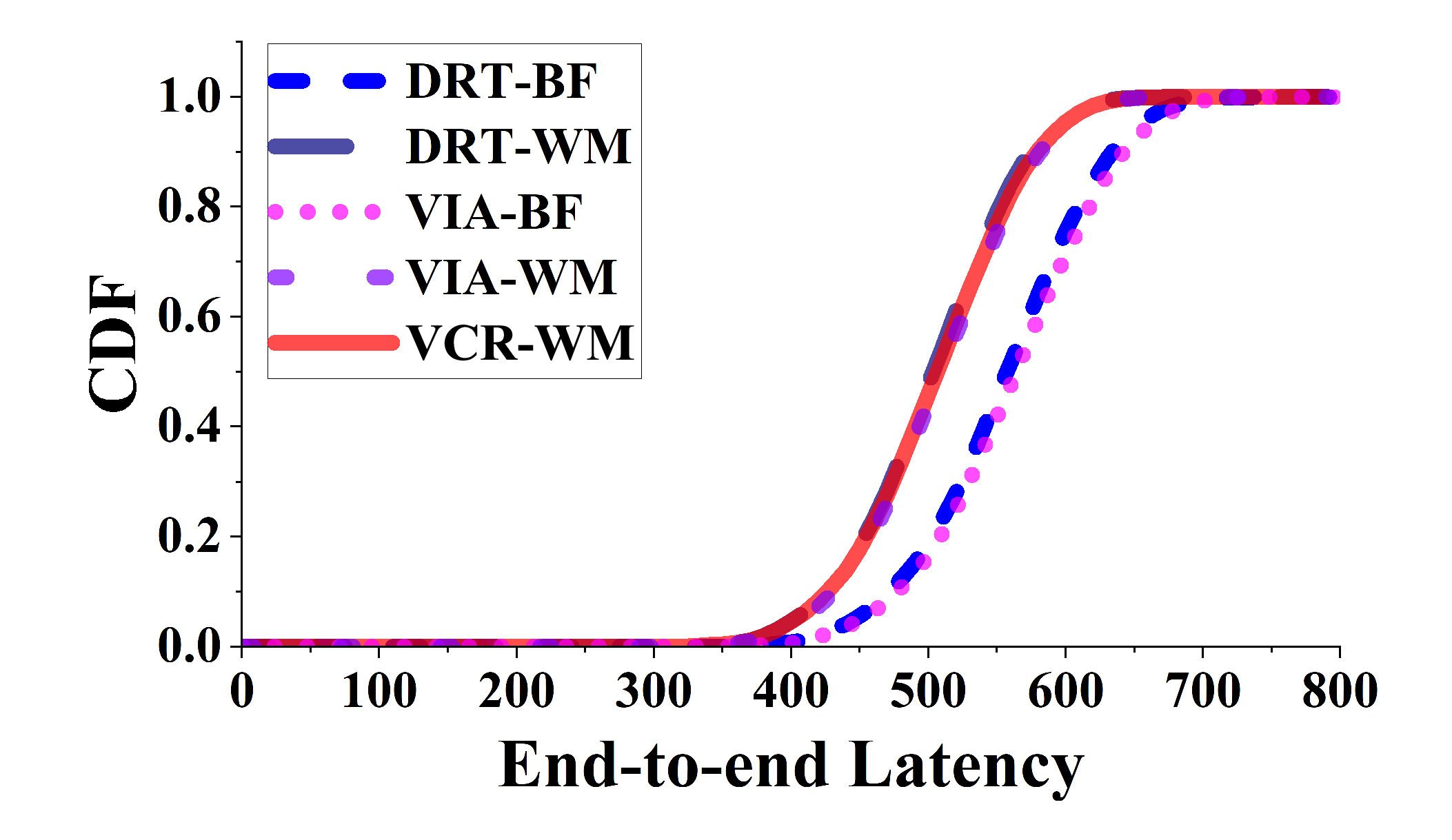}\caption{GR-UserUS}\label{fig:overall:gre_us}
	\end{subfigure}
 \hfil
    \begin{subfigure}{0.18\linewidth}
    \setlength{\abovecaptionskip}{-1pt}
	\hspace{-0.23cm}
	\includegraphics[width=1.2\linewidth]{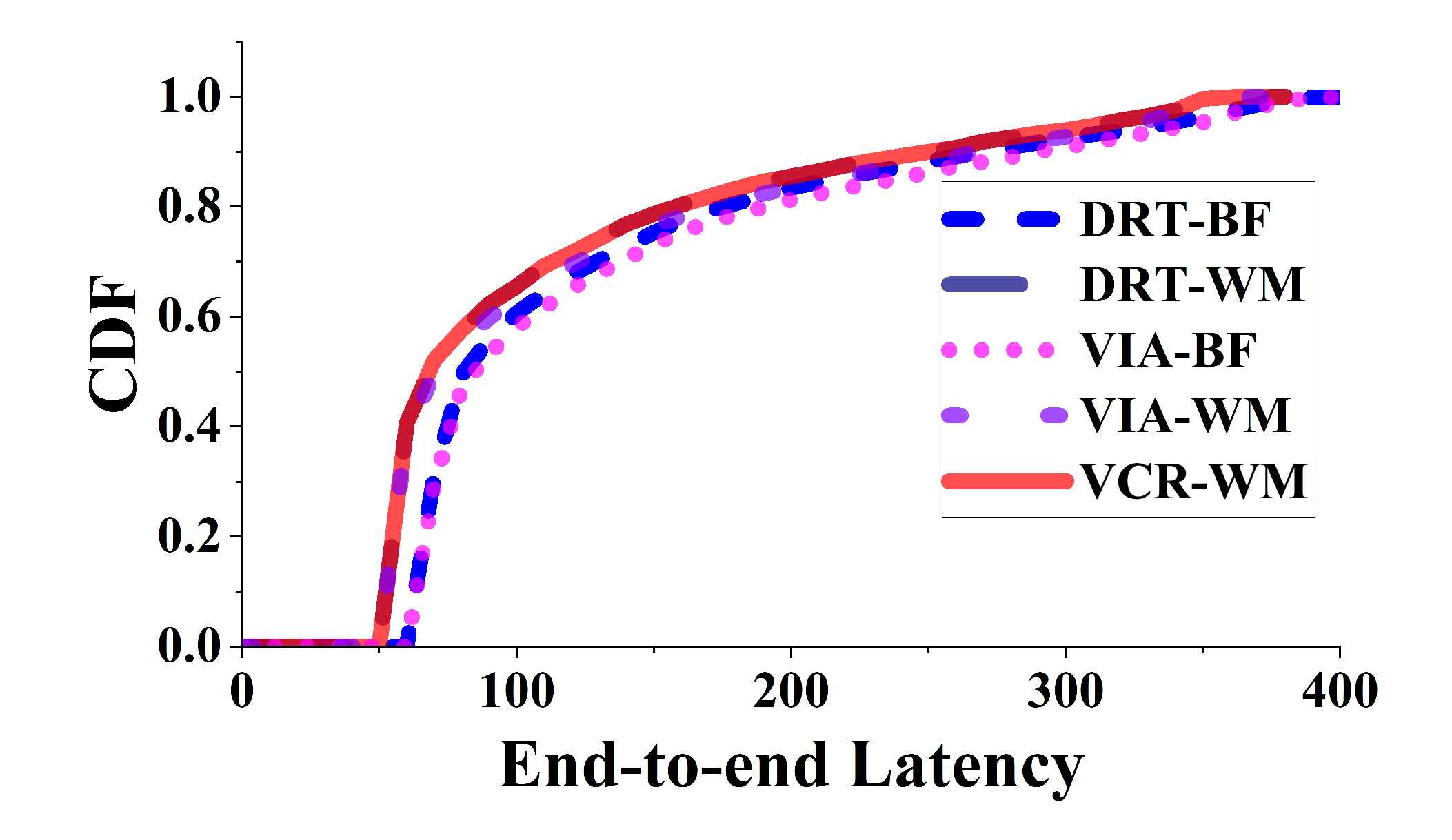}\caption{GR-UserLT}\label{fig:overall:gre_lith}
	\end{subfigure}
 \vspace{-1ex}
    \caption{End-to-end latency (ms) of a meeting session with endpoint located in Greece (WonderProxy). 
    }\label{fig:overall:gre}
\vspace{-2ex}\end{figure*}

\begin{figure*}[!htb]
    \centering
    \begin{subfigure}{0.18\linewidth}
    \setlength{\abovecaptionskip}{-1pt}
	\hspace{-0.23cm}
 	\includegraphics[width=1.2\linewidth]{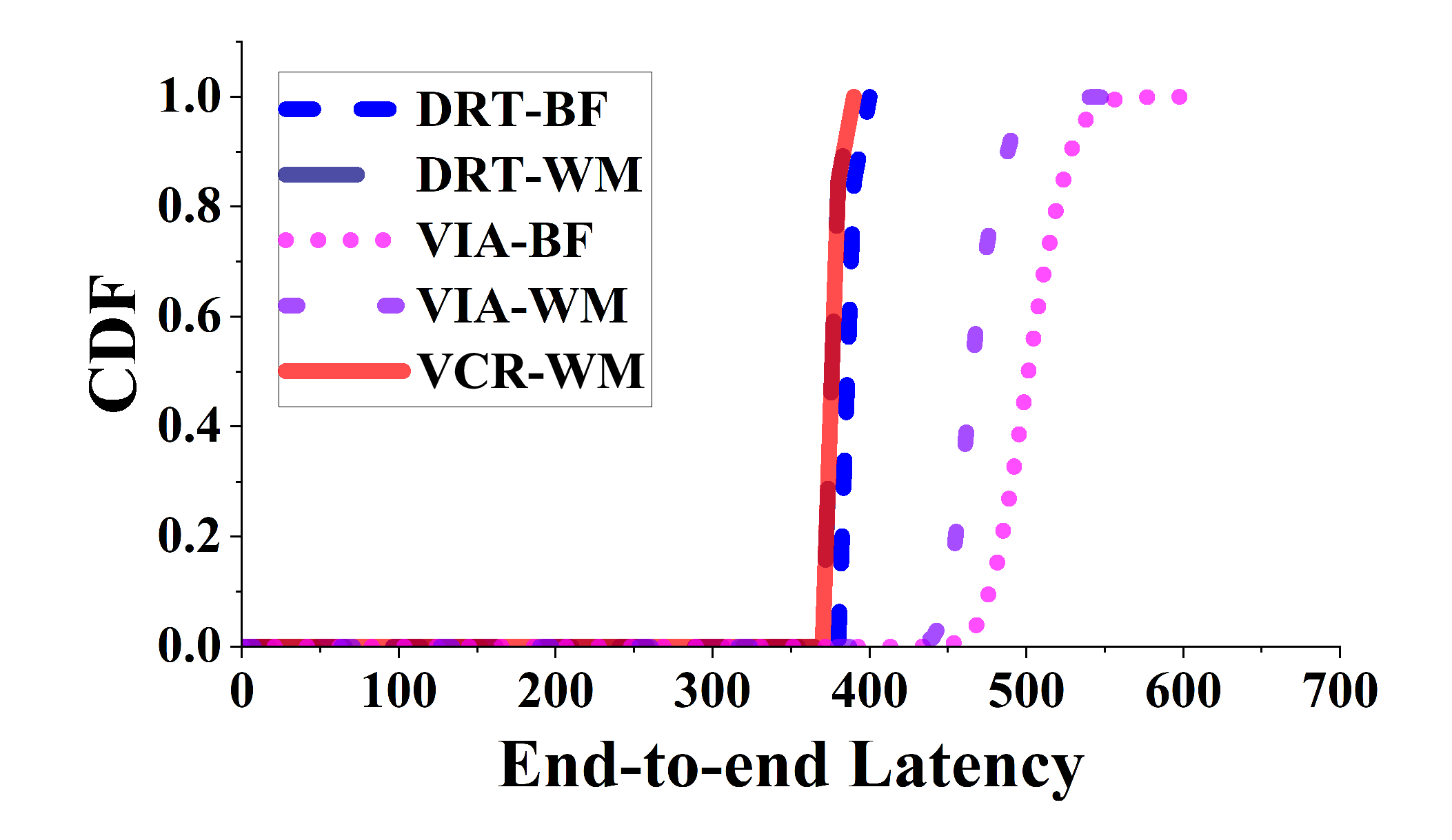}\caption{ZA-UserAR}\label{fig:overall:saf_arg}
	\end{subfigure}
 \hfil
    \begin{subfigure}{0.18\linewidth}
    \setlength{\abovecaptionskip}{-1pt}
	\hspace{-0.23cm}
	\includegraphics[width=1.2\linewidth]{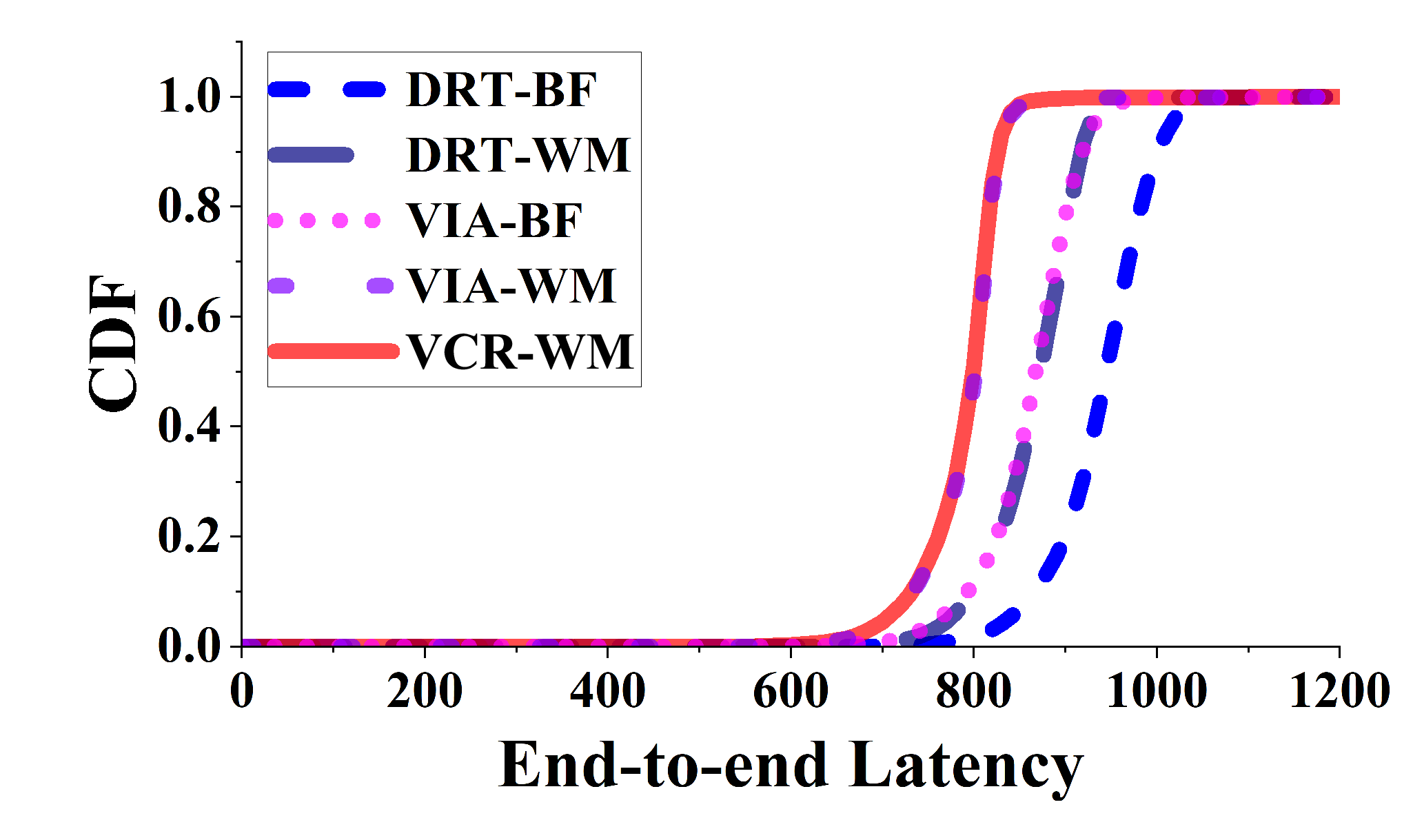}\caption{ZA-UserIDN}\label{fig:overall:saf_ind}
	\end{subfigure}
 \hfil
    \begin{subfigure}{0.18\linewidth}
    \setlength{\abovecaptionskip}{-1pt}
	\hspace{-0.23cm}
	\includegraphics[width=1.2\linewidth]{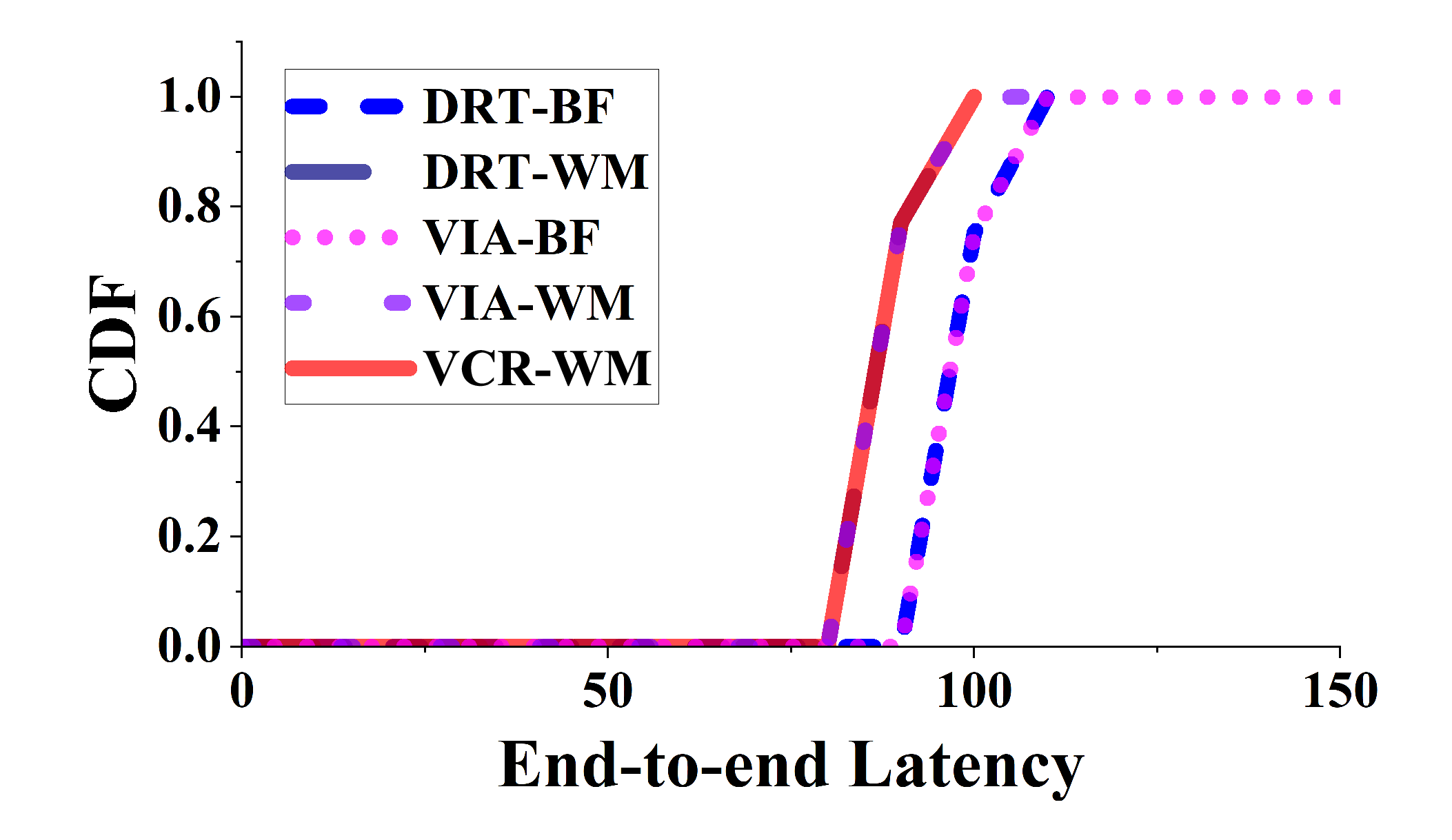}\caption{ZA-UserUGA}\label{fig:overall:saf_uga}
	\end{subfigure}
 \hfil
    \begin{subfigure}{0.18\linewidth}
    \setlength{\abovecaptionskip}{-1pt}
	\hspace{-0.23cm}
	\includegraphics[width=1.2\linewidth]{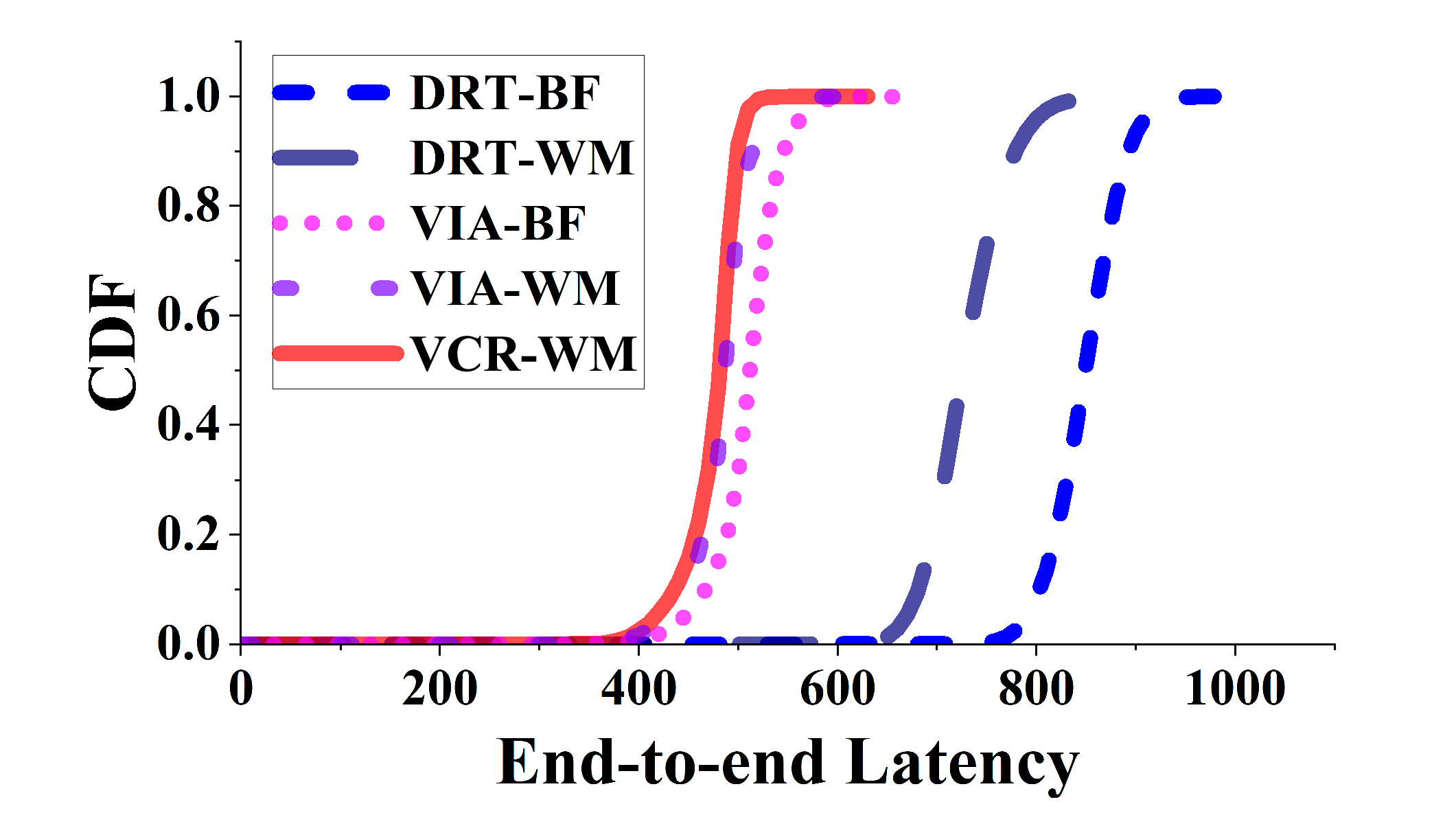}\caption{ZA-UserUS}\label{fig:overall:saf_us}
	\end{subfigure}
 \hfil
    \begin{subfigure}{0.18\linewidth}
    \setlength{\abovecaptionskip}{-1pt}
	\hspace{-0.23cm}
	\includegraphics[width=1.2\linewidth]{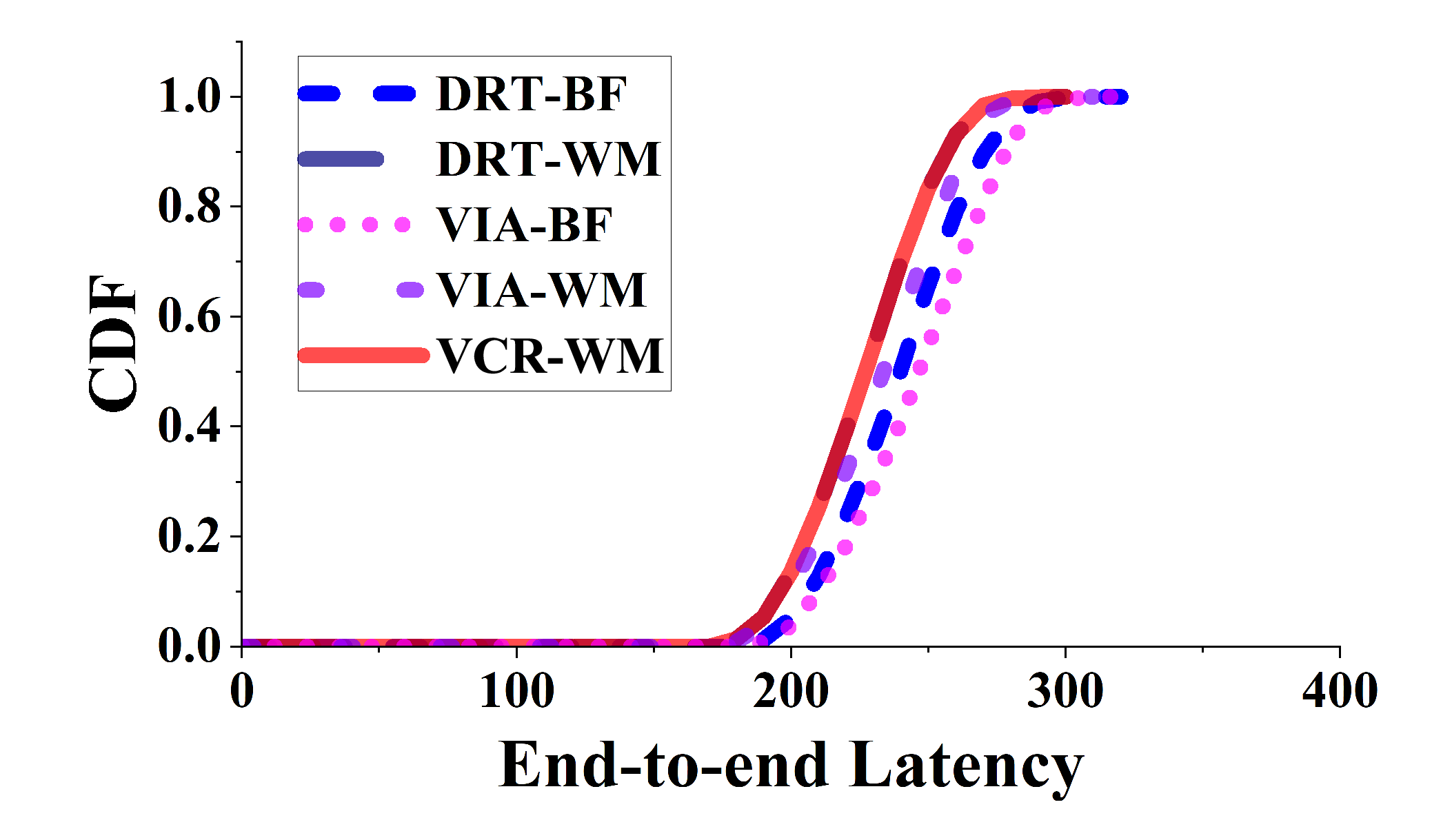}\caption{ZA-UserLT}\label{fig:overall:saf_lith}
	\end{subfigure}
 \vspace{-1ex}
    \caption{End-to-end latency (ms) of a meeting session with endpoint located in South Africa (WonderProxy). 
    }\label{fig:overall:aus}
\vspace{-2ex}
\end{figure*}

\begin{figure*}[!htb]
    \centering
    \begin{subfigure}{0.18\linewidth}
   \setlength{\abovecaptionskip}{-1pt}
	\hspace{-0.23cm}
 	\includegraphics[width=1.2\linewidth]{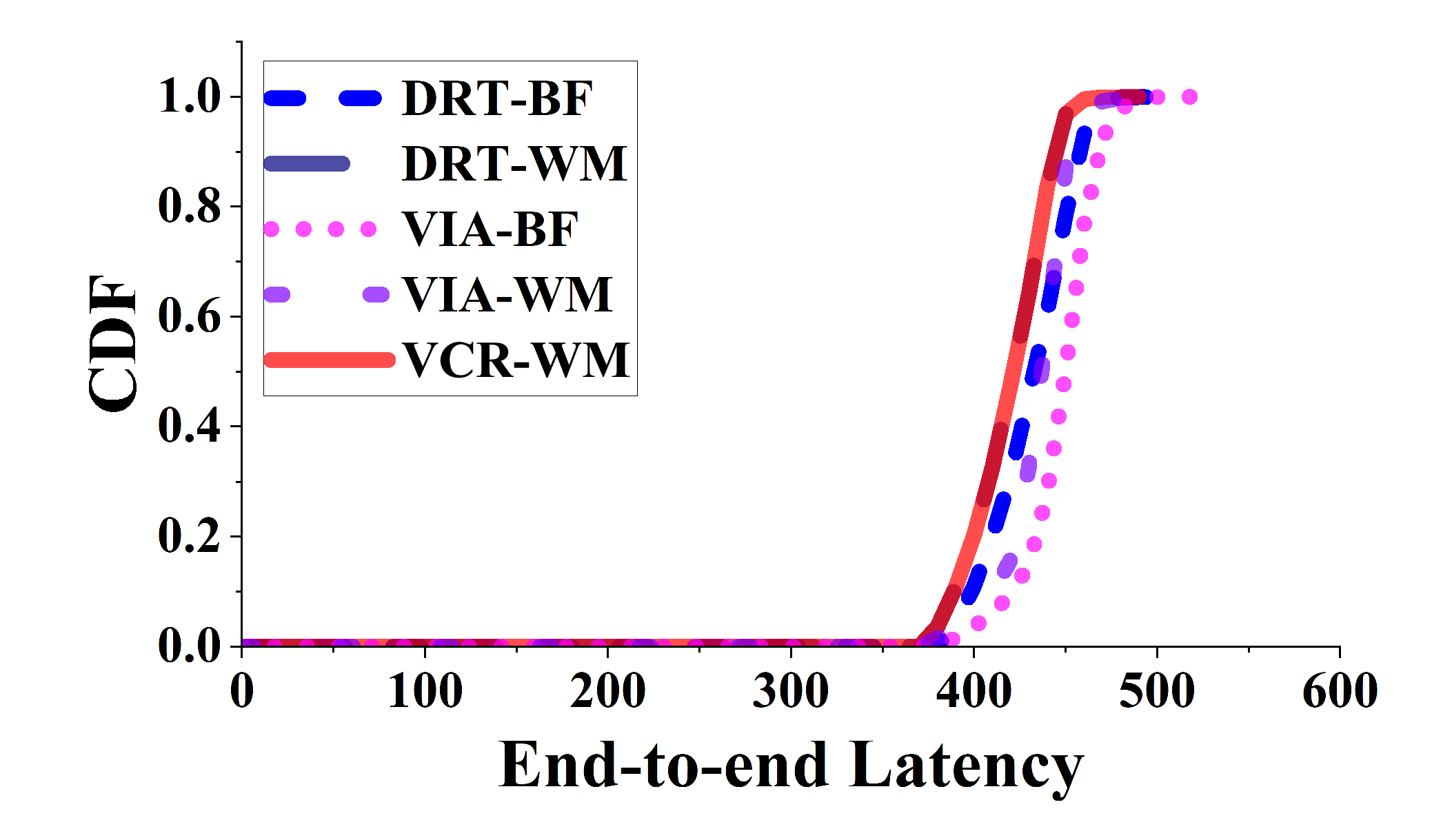}\caption{SG-UserAR}\label{fig:overall:sig_arg}
	\end{subfigure}
 \hfil
    \begin{subfigure}{0.18\linewidth}
    \setlength{\abovecaptionskip}{-1pt}
	\hspace{-0.23cm}
	\includegraphics[width=1.2\linewidth]{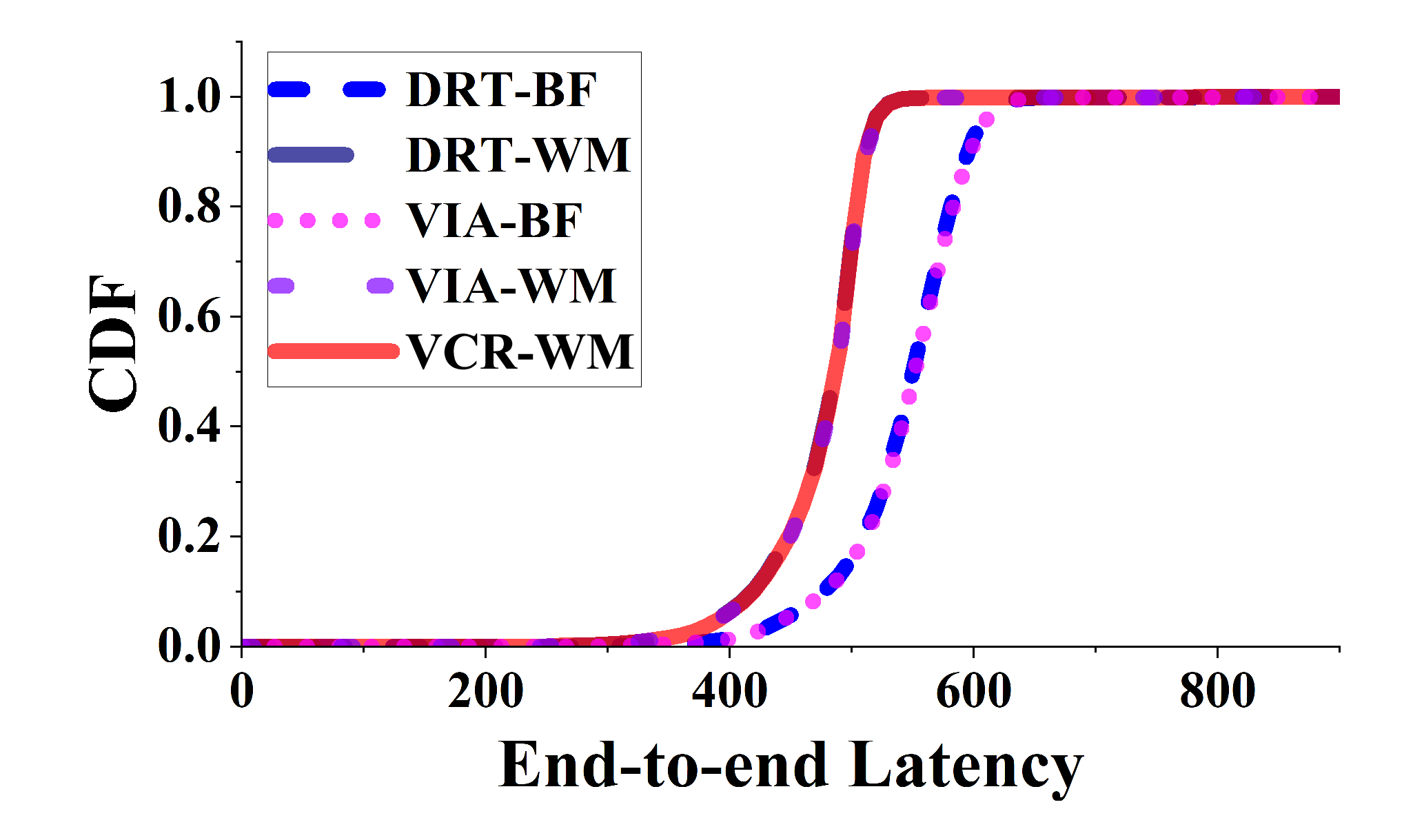}\caption{SG-UserIDN}\label{fig:overall:sig_ind}
	\end{subfigure}
 \hfil
    \begin{subfigure}{0.18\linewidth}
    \setlength{\abovecaptionskip}{-1pt}
	\hspace{-0.23cm}
	\includegraphics[width=1.2\linewidth]{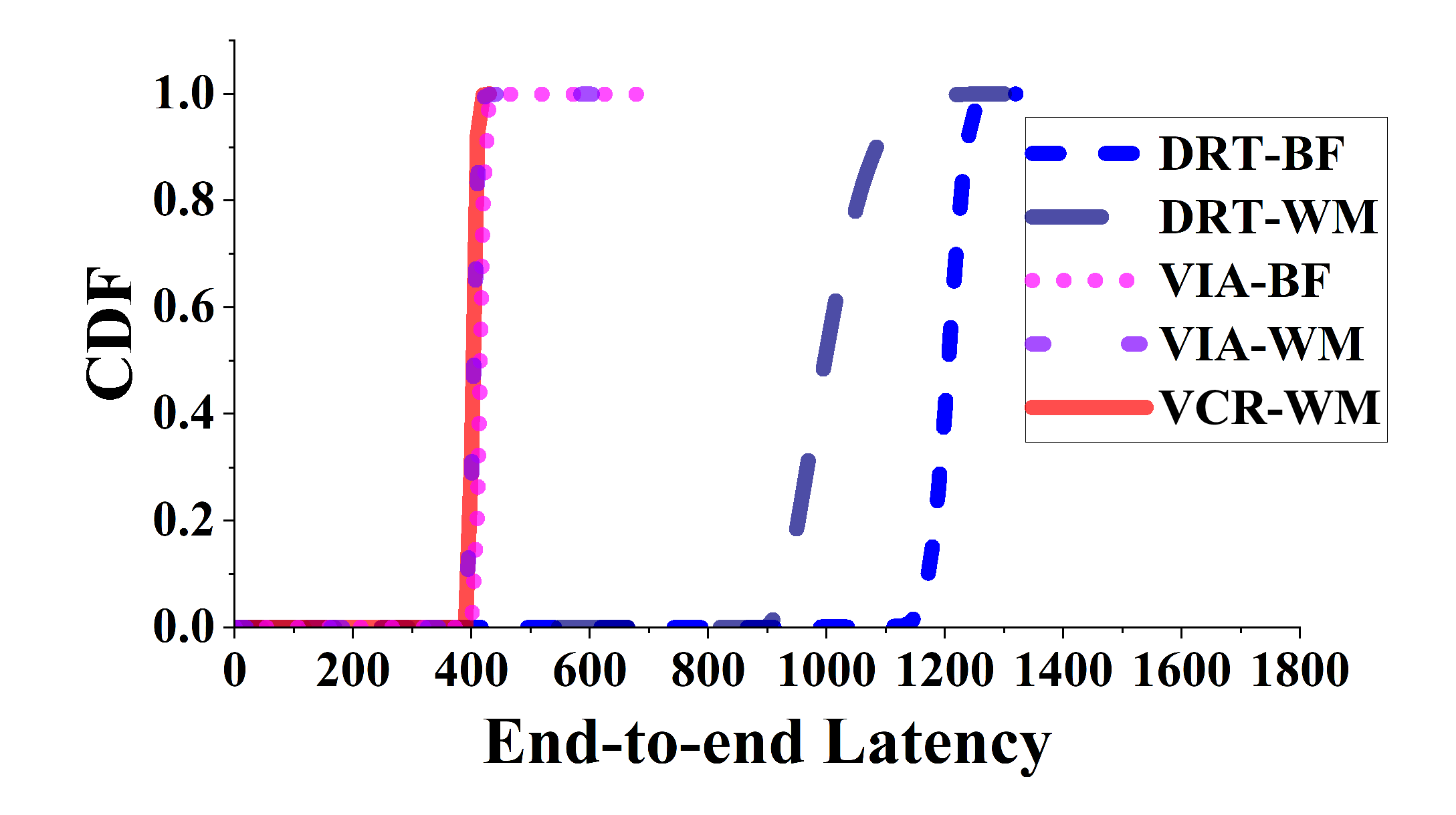}\caption{SG-UserUGA}\label{fig:overall:sig_uga}
	\end{subfigure}
 \hfil
    \begin{subfigure}{0.18\linewidth}
    \setlength{\abovecaptionskip}{-1pt}
	\hspace{-0.23cm}
	\includegraphics[width=1.2\linewidth]{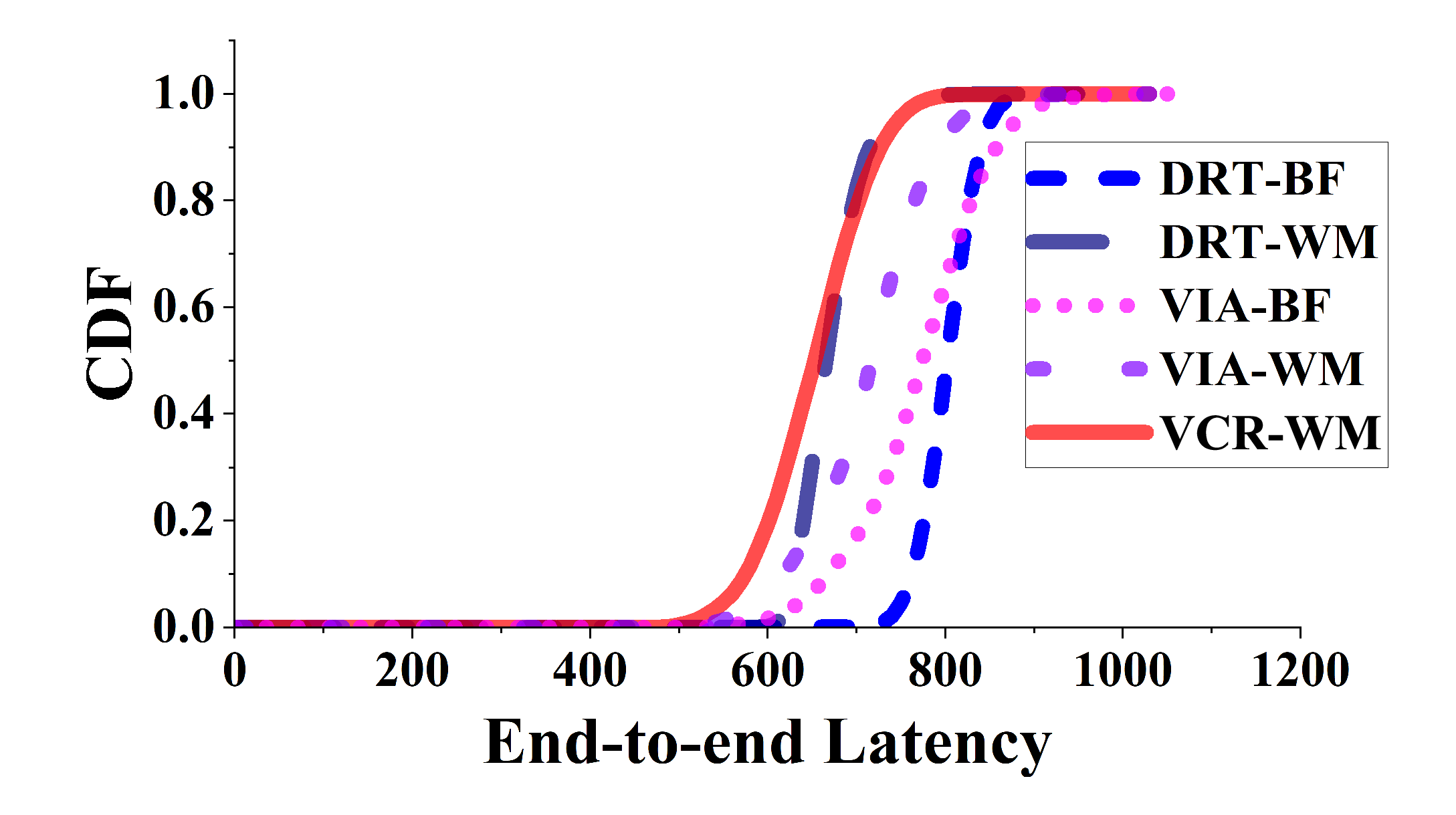}\caption{SG-UserUS}\label{fig:overall:sig_us}
	\end{subfigure}
 \hfil
    \begin{subfigure}{0.18\linewidth}
    \setlength{\abovecaptionskip}{-1pt}
	\hspace{-0.23cm}
	\includegraphics[width=1.2\linewidth]{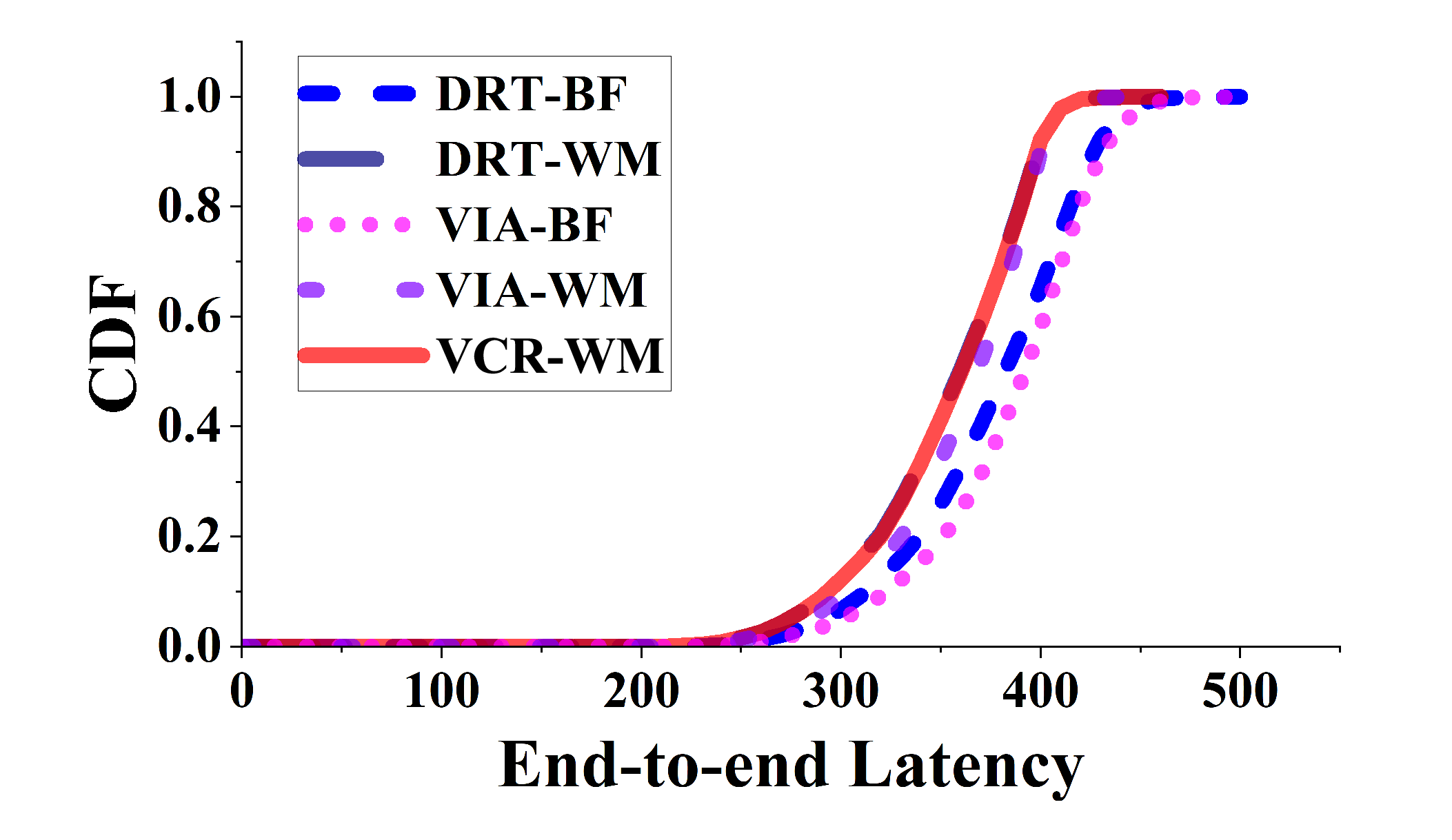}\caption{SG-UserLT}\label{fig:overall:sig_lith}
	\end{subfigure}
 \vspace{-1ex}
    \caption{End-to-end latency (ms) of a meeting session with endpoint located in Singapore (WonderProxy). 
    }\label{fig:overall:sig}
     \vspace{-2ex}
\end{figure*}

\subsection{Trace-Driven Evaluation Results}
\subsubsection{End-to-end latency on WonderProxy}\label{sec:eval:wonder}
To compare the end-to-end latency results obtained by different methods, we plot the CDFs of streaming lag experienced by clients in Figures~\ref{fig:overall:gre}-\ref{fig:overall:sig}. Results of meeting sessions hosted on the Brazil and Australia endpoints are similar to those shown in the figures, and thus are not plotted.
We have the following observations.

First, across all the meeting sessions hosted in different data centers, VCR-WM obtains the best end-to-end latency for all users distributed globally, while DRT-BF obtains the worst results in most cases. Take the results of Greece endpoint as an example. Figures~\ref{fig:overall:gre_uga} shows that, VCR-WM guarantees all packets to arrive within 640ms for users located in Uganda, which is even lower than the shortest end-to-end latency results obtained by DRT-BF. 
Compared to VIA-BF, VCR-WM reduces the average end-to-end latency by 4\%-13\%. 
Note that, for some cases, the VCR-WM and DRT-WM obtains similar end-to-end latency results (i.e., $<$ZA, UserAR$>$). This is because that the best relay path is the direct path, and the transmitting latency is far less than the other paths. Similarly, for some cases, the VCR-WM and VIA-WM obtains similar end-to-end latency results (i.e., $<$SG, UserUGA$>$) because the best path is far better than others.
Comparing different users in the same meeting session, users closer to the endpoint are more likely to obtain better video conferencing experiences. For example, Lithuania, which is also in Europe as Greece, is geographically the closest to the endpoint and thus has the lowest average end-to-end latency among all users.
What's more, all compared methods work similarly for Lithuania, since the direct route has much lower transmitting latency compared to the other relay routes. As a result, Direct, Via and VCRoute all select the same direct path. This further increases the stability of latency and reduces jitter.

Second, to study the effectiveness of WMJitter, we compare the results of DRT-BF with those of DRT-WM and the results of VIA-BF with those of VIA-WM.
As shown in Figure~\ref{fig:net:wonder}, the network performance between endpoints and users IDN, UGA and US are much more dynamic than the other two users. As a result, we have observed much higher improvement of the watermark-based jitter management methods compared to the buffer-based method on reducing the end-to-end latency. For example, DRT-WM reduces the end-to-end latency by 8\%-12\%, 9\%-16\% and 10\%-16\% compared to DRT-BF for clients in IDN, UGA and US, respectively, across the three meeting sessions.
Similarly, VIA-WM reduces the end-to-end latency by 8\%-13\%, 2\%-9\% and 5\%-10\% compared to VIA-BF for clients in IDN, UGA and US, respectively, across the three meeting sessions.
The above results have demonstrated the effectiveness of WMJitter.

\subsubsection{End-to-end latency on Tencent Cloud}\label{sec:eval:tencent}
Evaluation results on Tencent cloud has shown similar observations as those from WonderProxy. Below we only discuss the differences between Tencent cloud and WonderProxy.

\begin{figure}[!htb]
	\centering
	\begin{subfigure}{0.45\linewidth}
		\setlength{\abovecaptionskip}{-1pt}
		\includegraphics[width=1.1\linewidth]{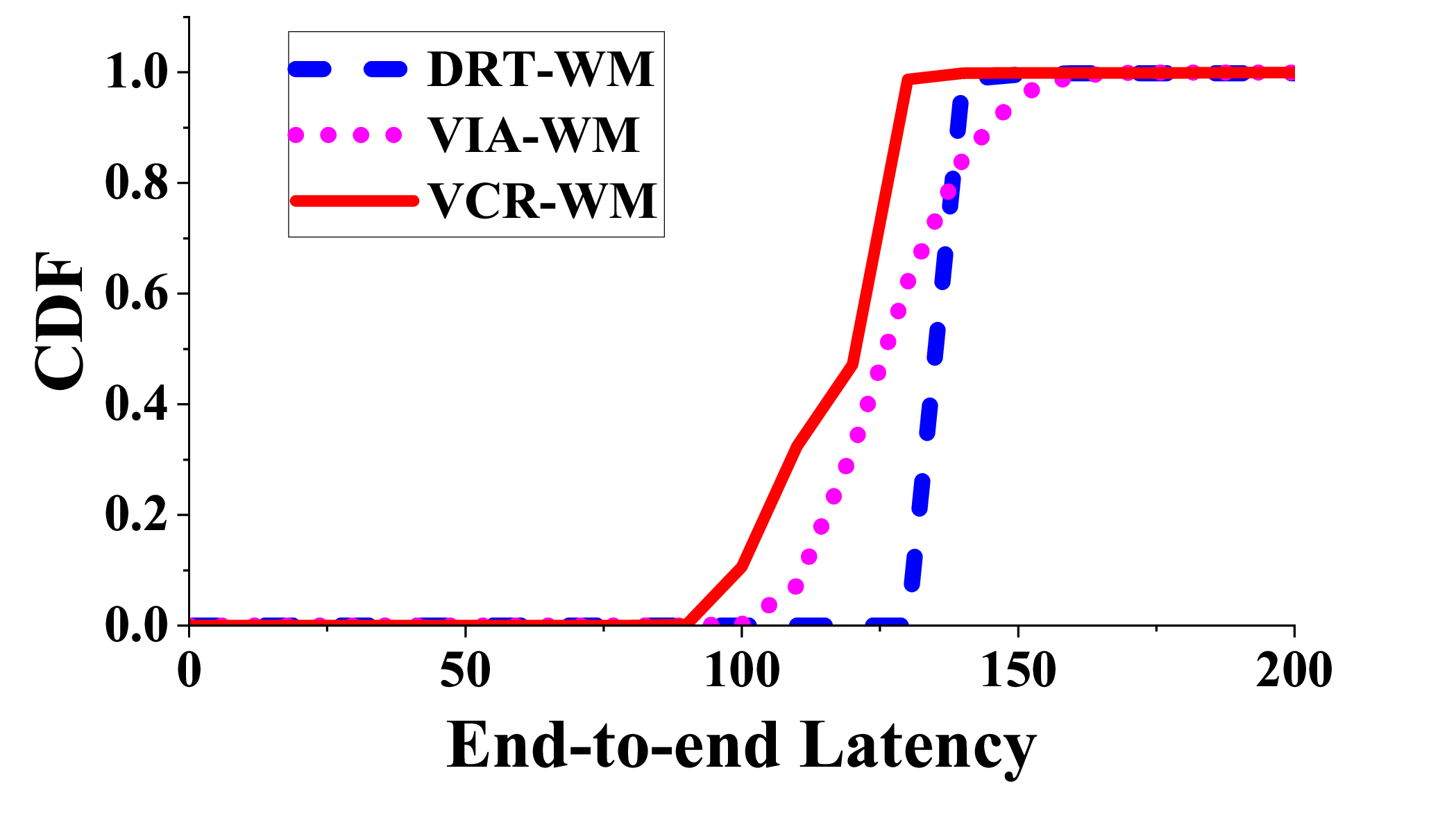}
		\caption{TYO-UserSZ}\label{fig:overall:dj_xuexiao}
	\end{subfigure}
	\hfill
	\begin{subfigure}{0.45\linewidth}
		\setlength{\abovecaptionskip}{-1pt}
		\includegraphics[width=1.1\linewidth]{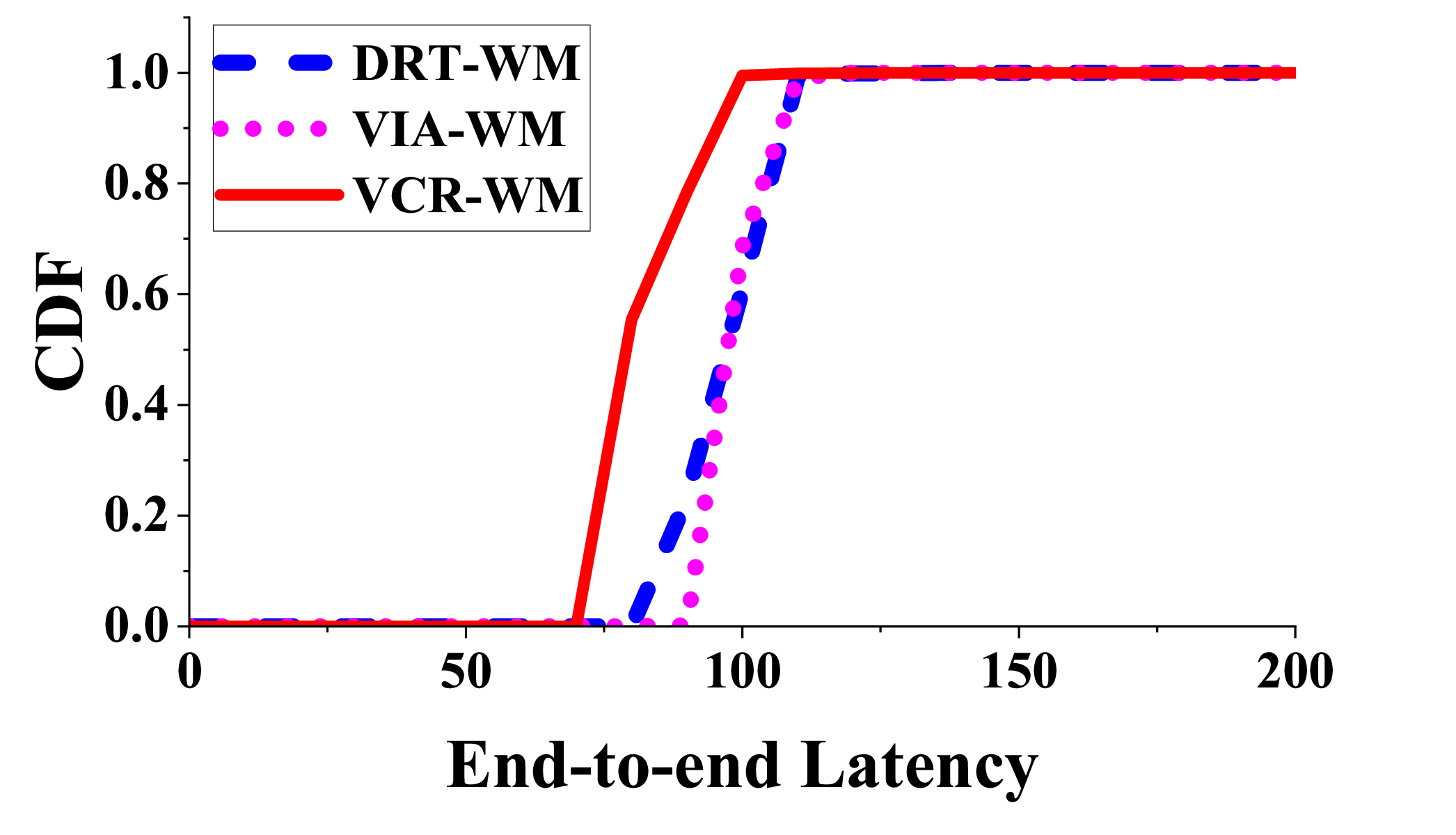}\caption{TYO-UserSG}\label{fig:overall:dj_xjploc}
	\end{subfigure}
	\vspace{-1ex}
	\caption{End-to-end latency (ms) of a meeting session with endpoint located in Tokyo (Tencent).
	}\label{fig:overall:dj}
	\vspace{-2ex}
\end{figure}

\begin{figure}[!htb]
	\centering
	\begin{subfigure}{0.45\linewidth}
		\setlength{\abovecaptionskip}{-1pt}
		\includegraphics[width=1.1\linewidth]{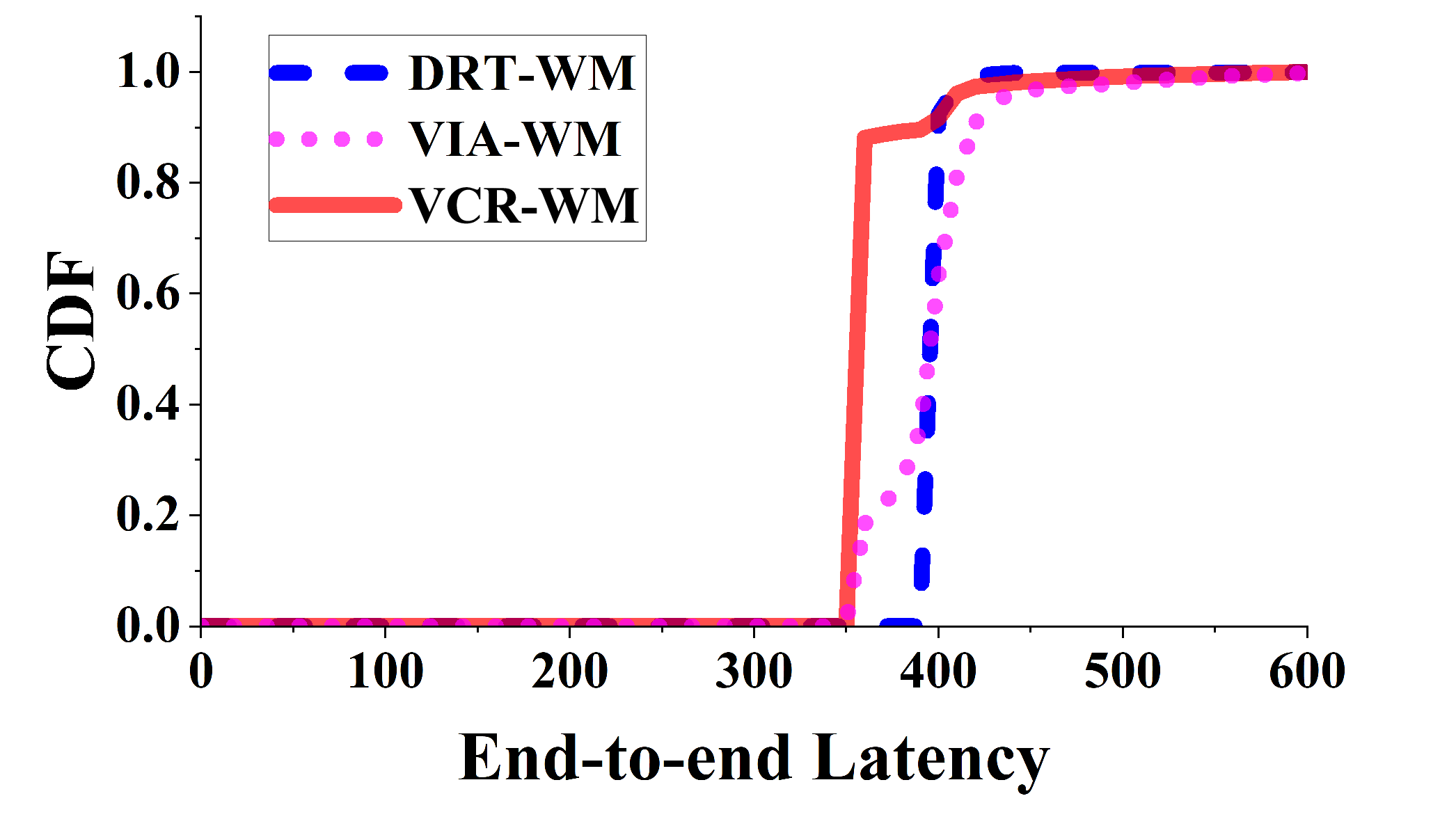}\caption{SP-UserSZ}\label{fig:overall:sbl_xuexiao}
	\end{subfigure}
	\hfill
	\begin{subfigure}{0.45\linewidth}
		\setlength{\abovecaptionskip}{-1pt}
		\includegraphics[width=1.1\linewidth]{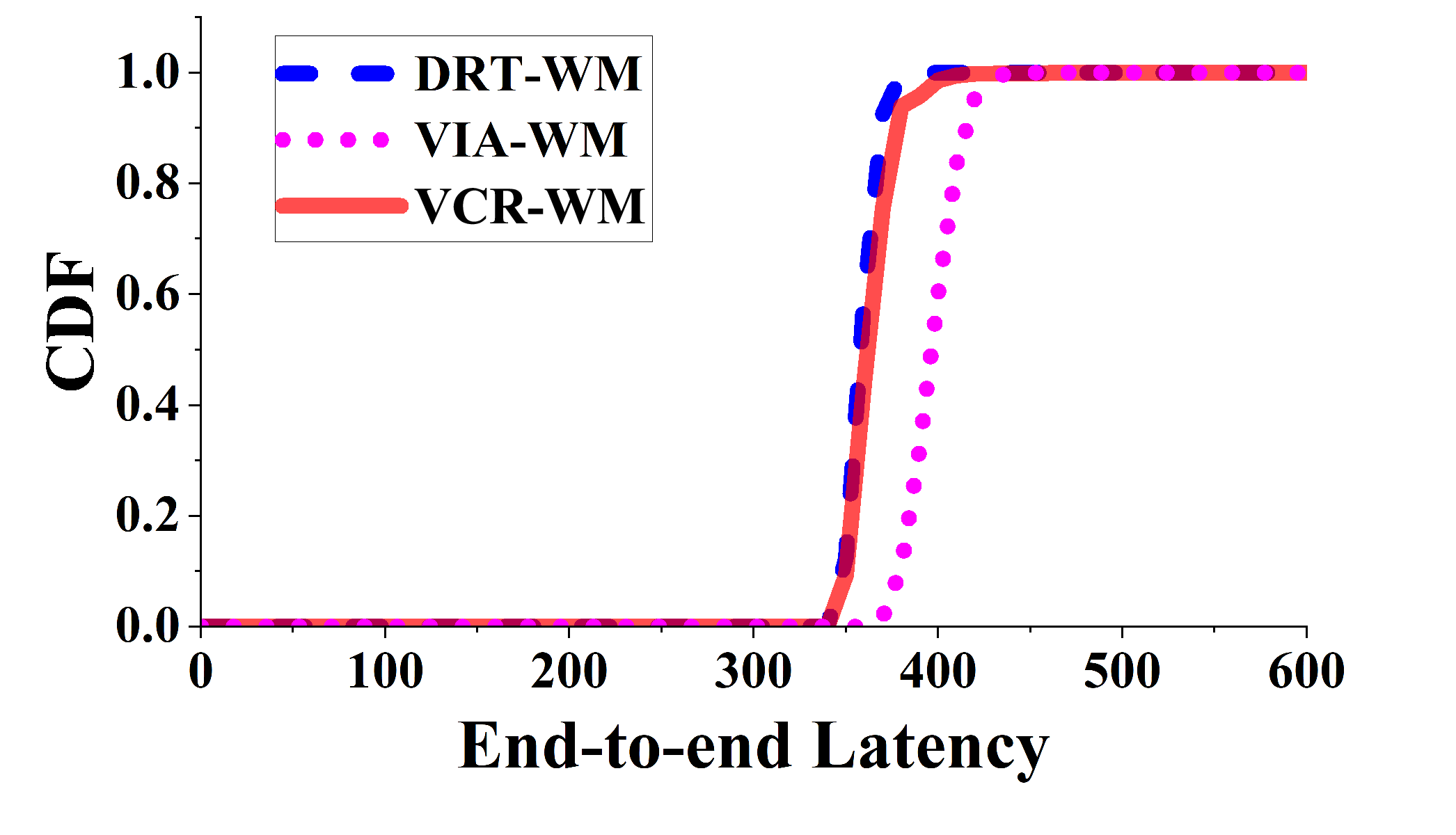}\caption{SP-UserSG}\label{fig:overall:sbl_xjploc}
	\end{subfigure}
	\vspace{-1ex}
	\caption{End-to-end latency (ms) of a meeting session with endpoint located in Sao Paulo (Tencent). 
	}\label{fig:overall:sbl}
	\vspace{-2ex}
\end{figure}

Figure~\ref{fig:overall:dj} and \ref{fig:overall:sbl} show the CDFs of end-to-end latency results of users obtained by VCR-WM, DRT-WM and VIA-WM when meetings are hosted on the Tokyo and Sao Paulo servers. We deliberately removed the results of DRT-BF and VIA-BF from the figures since they performed very closely to DRT-WM and VIA-WM due to the low variances of the network performances.
On these two cases, VCR-WM obtains much lower end-to-end latency compared to VIA-WM and DRT-WM. For example, the 50\%-percentile latency of VCR-WM is 0\%-16\% and 6\%-18\% lower than those of DRT-WM and VIA-WM, respectively.
This demonstrates the effectiveness of VCRoute on finding better routes than other state-of-the-art routing schedulers for video conferencing systems.


\subsubsection{Loss rate results}
\begin{figure}[t]
	\centering
	\begin{subfigure}{0.45\linewidth}
		\setlength{\abovecaptionskip}{-1pt}
		\hspace{-0.55cm}
		\includegraphics[width=1.1\linewidth]{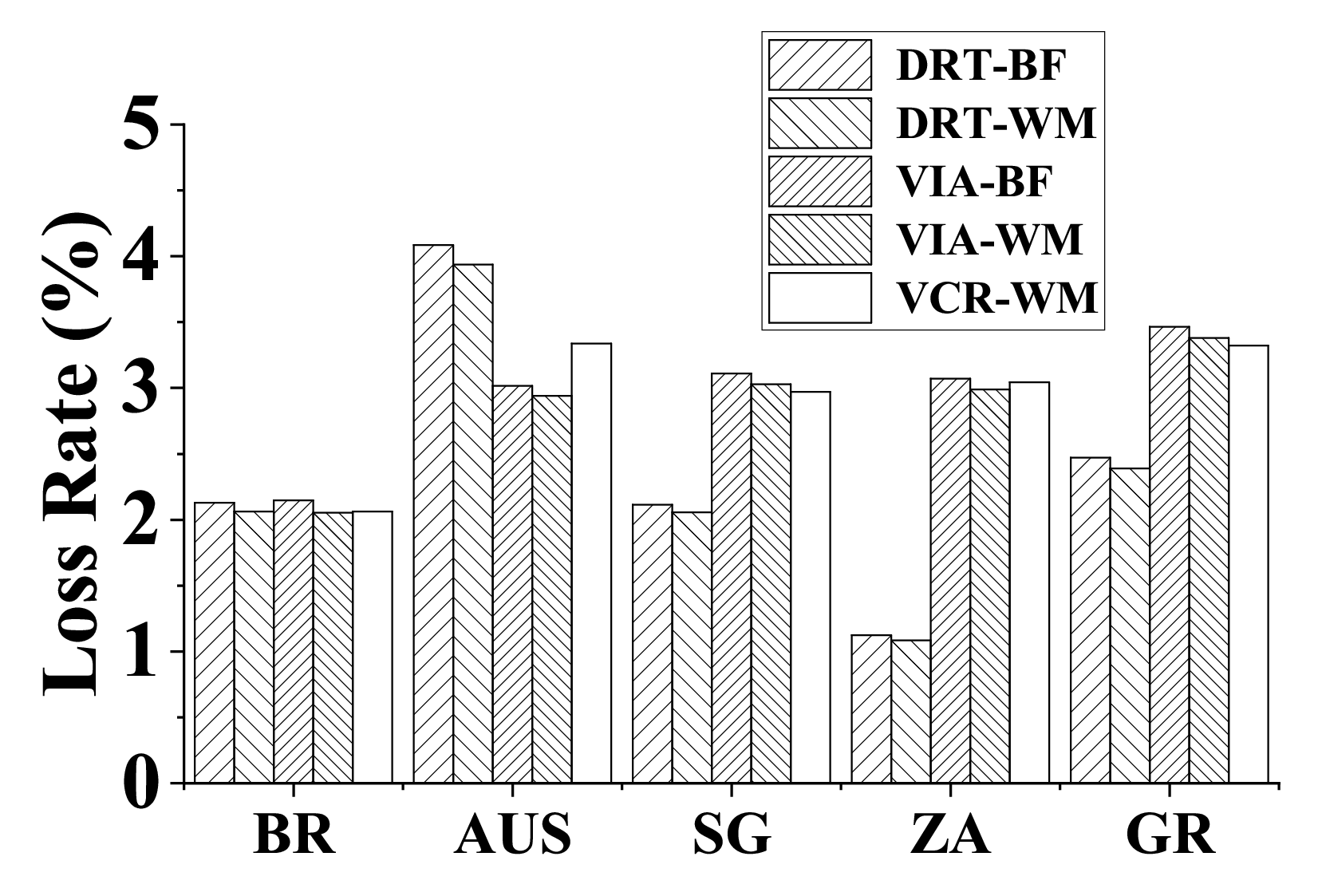}
		\caption{ WonderProxy}\label{fig:overall:lossrate_wonder}
	\end{subfigure}
	\hfill
	\begin{subfigure}{0.45\linewidth}
		\setlength{\abovecaptionskip}{-1pt}
		\hspace{-0.55cm}
		\includegraphics[width=1.1\linewidth]{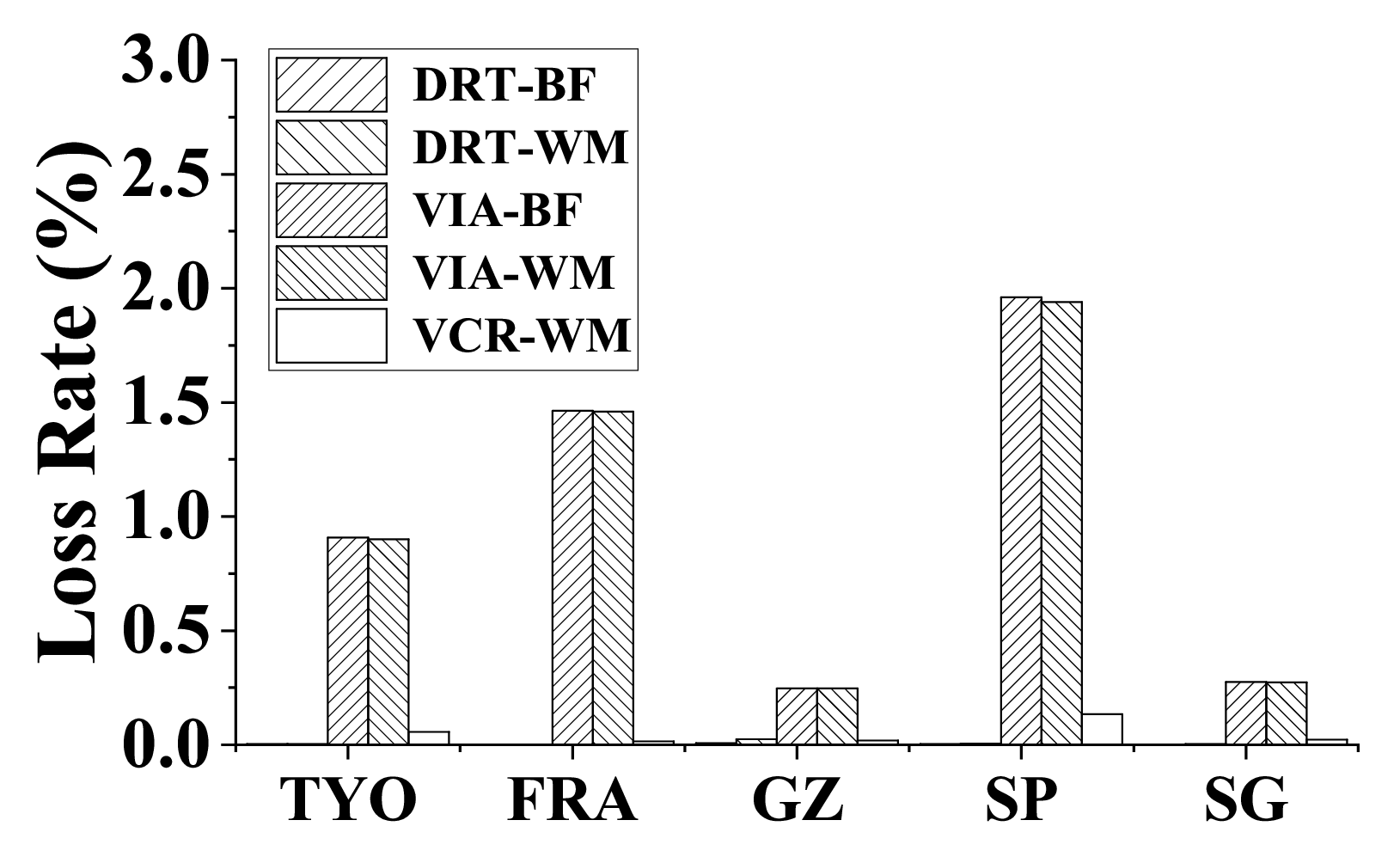}\caption{Tencent cloud}\label{fig:overall:lossrate_tencent}
	\end{subfigure}
	\vspace{-1ex}
	\caption{Overall loss rate of meeting sessions hosted on different endpoints in WonderProxy and Tencent. 
	}\label{fig:overall:lossrate}
	\vspace{-2ex}
\end{figure}

We further study the loss rate of audio streams resulted from different comparisons.
Figure~\ref{fig:overall:lossrate} shows the loss rate results per meeting session on both WonderProxy and Tencent cloud. As expected, the loss rate on WonderProxy is in general much larger than that on Tencent cloud, due to the much higher network performance variances on WonderProxy.
The loss rates of meeting sessions using Direct routing on Tencent cloud are almost zero since there's no significant jitter on the route. Thus no packet arrives too late to be dropped. In contrary, the loss rate is high even on direct routes on WonderProxy due to the high network performance variances. Further, watermark-based jitter management method can slightly reduce the loss rate compared to buffer-based methods. For example, DRT-WM and VIA-WM reduce the loss rate by 2\%-3\% and 2\%-4\% compared to DRT-BF and VIA-BF, respectively, on WonderProxy. 
The Via routing method (i.e., VIA-BF and VIA-WM) generated much higher loss rates on Tencent cloud than the other methods, due to its tendency to explore which leads to more path changes. For example, there are 34,020 times of path changes with Via for the $<$Tokyo, Shenzhen$>$ data stream with 600,000 packets.
Although VCRoute also explores during the search for a good route, it deliberately reduces the number of path changes to reduce network jitter.
In all cases, VCR-WM yields lower or comparable packet loss rates compared to the other methods, while providing much lower end-to-end latency.

\subsubsection{System overhead}
\begin{figure}[t]
    \centering
    \begin{subfigure}{0.4\linewidth}
    \setlength{\abovecaptionskip}{-1pt}
	\hspace{-0.55cm}
 	\includegraphics[width=1.1\linewidth]{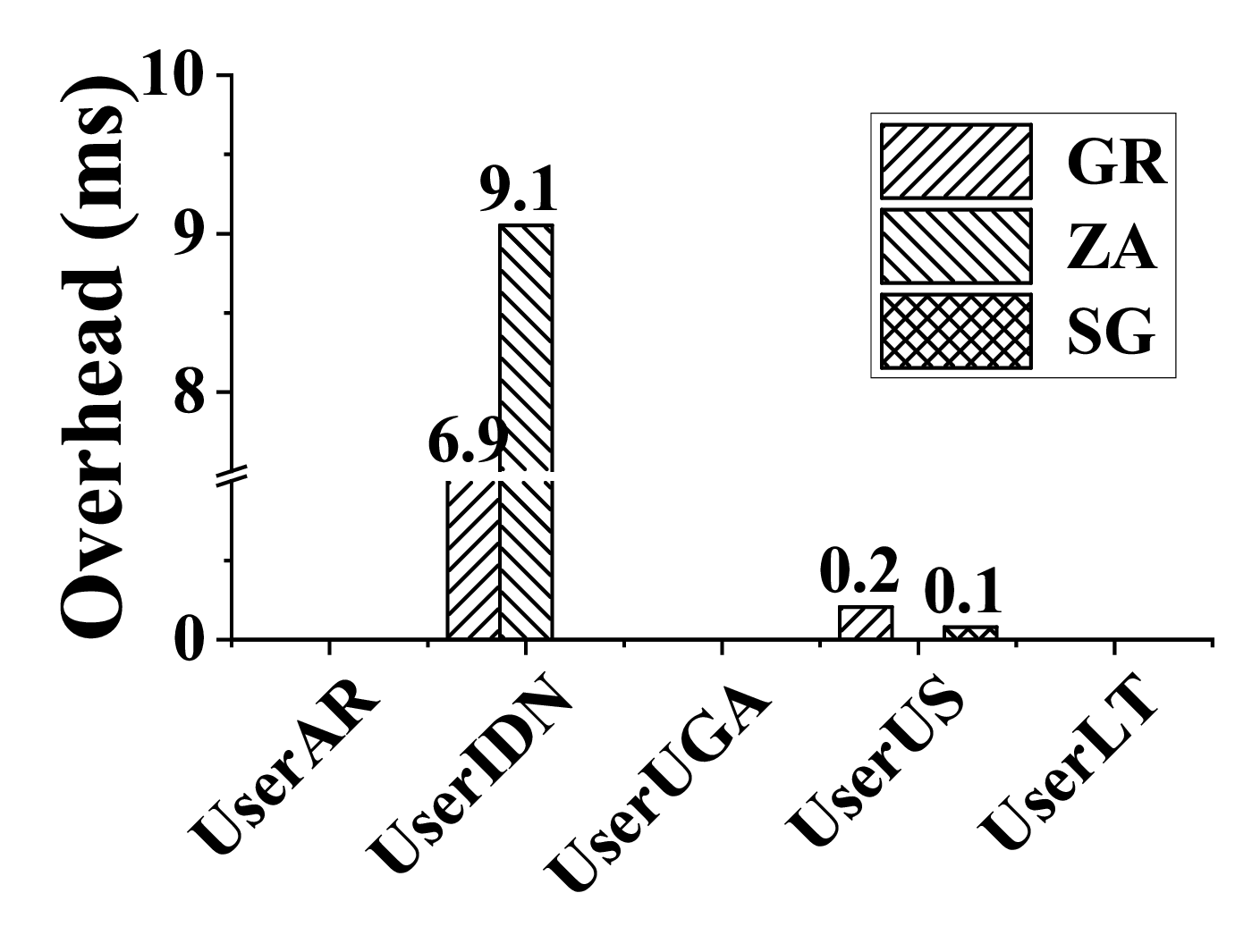}\caption{ WonderProxy}\label{fig:overhead:wonder}
	\end{subfigure}
 \hfil
    \begin{subfigure}{0.4\linewidth}
    \setlength{\abovecaptionskip}{-1pt}
	\hspace{-0.55cm}
	\includegraphics[width=1.1\linewidth]{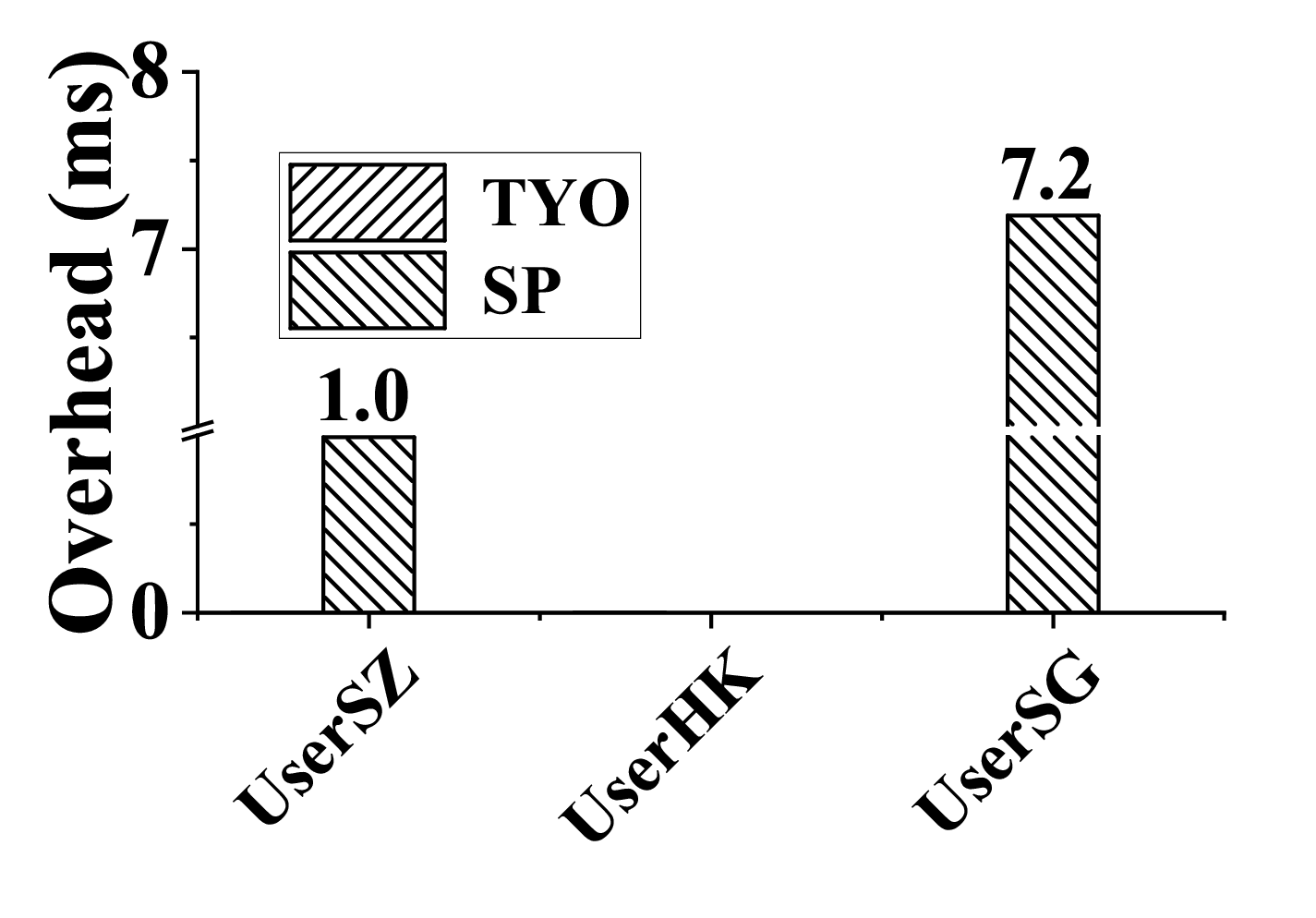}\caption{Tencent cloud}\label{fig:overhead:tencent}
	\end{subfigure}
 \vspace{-1ex}
    \caption{Additional overhead per packet of VCRoute}\label{fig:overhead}
     \vspace{-2ex}
\end{figure}

With our top-k pruning, the overhead of VCRoute searching for the best route is very low ($\sim$0.3ms for each packet). Due to the centralized design of VCRoute, the main overhead introduced by our system to audio streams comes from the extra time needed to transfer routing decisions from endpoints to users. Figure~\ref{fig:overhead} shows the additional overhead per packet caused by transmitting routing decisions for each meeting session hosted on different endpoints in WonderProxy and Tencent.
Similar to our observations above, the additional overhead is larger for users with higher network variances (e.g., UserIDN on WonderProxy and UserSG on Tencent). The largest overhead for UserIDN is 9.1ms, which is neglectable compared to the end-to-end latency per packet as shown in Figure~\ref{fig:overall:saf_ind}. It means that our design introduces an additional overhead less than 1\% of the average end-to-end latency for each packet.

\vspace{-1ex}
\subsection{Discussions}
The above results show that our proposed methods are effective and efficient to different geo-distributed environments. 
\begin{itemize}[leftmargin=*]
    \item VCRoute is effective for DCs with highly heterogeneous network, as demonstrated by the results on Tencent cloud. 
    \item The WM-based jitter manager is effective for highly dynamic network environments (i.e., high jitter) as demonstrated by the results on WonderProxy.
    \item Our methods introduces an additional overhead less than 1\% of the average end-to-end latency and can scale well with the increase of network sizes.
    \vspace{-1ex}
\end{itemize}

\section{Related Work}\label{sec:related}

\subsection{Conferencing Systems}
Wu et al. \cite{vskyconf} presented a conferencing system vSkyConf, which use decentralized routing algorithm to reduce streaming latency. vSkyConf also adapt streaming rate according to the bandwidth limitation, and utilize buffer fro streaming synchronization. 
Fouladi et al. \cite{salsify} proposed to achieve low-latency video streaming by integrating video codec and transport protocol. 
Yan et al. \cite{situ} used a neural network to predict the transfer time of a chunk in a video, which is used for decision of bit rate selection for video streaming.
Fang et al. \cite{reinforce} also choose to control the sending rate of video streaming based on the bandwidth evaluation on the receiver. It used a RL-based method to change the sending rate.
Zhang et al. \cite{OnRL} also proposed to adapt transport protocol and video codec based on network dynamics. It proposed to train a RL-based model online, and used it to predict network dynamics.
Chang et al. \cite{10.1145/3487552.3487847} presented a detailed measurement study which compares Zoom, Webex and Google Meet. They found that users in different regions can have big difference streaming lag.
Hu et al. \cite{dejavu} proposed to enhancing video quality at the endpoint without changing the transmitting condition of the video.
Although these studies shed some lights on how to model video conferencing systems, none of them have addressed the latency issues studied in this paper.

\vspace{-1ex}
\subsection{Data Stream Processing}
Data stream processing research has delved into parallelism, synchronization, and resource optimization. 
Zhang et al. \cite{zhang2019briskstream} introduced BriskStream, a NUMA-aware in-memory system for shared-memory multicore architectures. 
Mencagli et al. \cite{mencagli2017parallel} developed a framework for parallel continuous preference queries on out-of-order, bursty data streams. 
Traub et al. \cite{traub2018scotty}  presented Scotty, a high-throughput operator for streaming window aggregation, while Miao et al. \cite{miao2017streambox} established StreamBox, an engine for out-of-order record processing on multicore servers. 
Zhang et al. \cite{zhang2020towards} developed TStream, supporting efficient concurrent state access with dynamic restructuring execution. 
Distinguishing our work, we introduce watermark-based jitter management, departing from traditional buffer-based approaches, enabling more efficient data processing and further reducing latency in geo-distributed environments.

\vspace{-1ex}
\subsection{Geo-distributed Stream Processing}
Network optimizations for geo-distributed stream processing have aimed to minimize communication costs, balance resource utilization, and adapt to bandwidth capacities. 
Gu et al. \cite{gu2015general} developed a framework for communication cost minimization using VM placement and flow balancing. 
Rabkin et al. \cite{rabkin2014aggregation} created JetStream, a wide-area data analysis system addressing bandwidth limitations. 
Zhang et al. \cite{zhang2018awstream} presented AWStream, offering low latency and high accuracy in bandwidth-constrained operations. 
Liu et al. \cite{liu2013bellini} designed Bellini, a system for rapid prototyping of inter-datacenter protocols and efficient VM resource utilization. 
Xu and Li \cite{xu2013joint} optimized joint request mapping and response routing for distributed data centers. 
Zhao et al. \cite{zhao2019optimizing} designed HPS+, a coordinated task scheduling and routing system. 
Mostafaei et al. \cite{mostafaei2021snr} devised SNR, a worker node placement approach balancing bandwidth, latency, and cost. 
Li et al. \cite{li2018wide} extended Spark Streaming for optimal task scheduling and data flow routing. 
Zoom \cite{Zoom} ensures optimized connection paths through geo-distributed infrastructure. Our work advance state-of-the-art through application-aware network routing and watermark-based low-latency jitter management, offering new strategies for optimizing low-latency systems, focusing on video conferencing applications in geo-distributed environments.

\vspace{-1ex}
\section{Conclusion}\label{sec:conclude}

In this paper, we address two challenges for video conferencing systems across geo-distributed DCs. First, existing routing methods mainly focus on minimizing packet transmitting latency, which does not necessarily translate to low end-to-end latency. In response, we introduced VCRoute, an application-specific packet routing technique that jointly considers network transmission time and packet reordering time to reduce end-to-end latency. Second, inter-DC network latency can drastically fluctuate, resulting in high jitter detrimental to video conferencing applications. Traditional buffer-based jitter management methods often introduce unnecessary delays, especially when handling stragglers. To overcome this, we proposed WMJitter, a watermark-based Out-of-Order Processing mechanism tailored to manage jitter at the users' end. Evaluations using two distinct real-world geo-distributed environments have demonstrated the efficacy and viability of our proposed solutions on enhancing the performance of video conferencing systems.

\vspace{-2ex}

\bibliographystyle{IEEEtran}
\bibliography{streaming}

\end{document}